\documentclass[useAMS,usenatbib]{mn2e}

\usepackage{graphicx}
\usepackage{ amsmath }
\usepackage{booktabs}
\usepackage{rotating}
\usepackage{morefloats}
\usepackage{ amssymb }

\title[Formation and X-ray Emission from Hot bubbles in PNe
(I)]{Formation and X-ray Emission from Hot bubbles in Planetary
  Nebulae\\ {\LARGE I. Hot bubble formation}}

\author[Toal\'{a} \& Arthur]{J.A.\,Toal\'{a}$^{1}$\thanks{E-mail:
    toala@iaa.es} and S.J.\,Arthur$^{2}$\\ $^{1}$Instituto de
  Astrof\'{\i}sica de Andaluc\'{\i}a, IAA-CSIC, Glorieta de la
  Astronom\'{\i}a s/n, 18008 Granada, Spain,\\ $^{2}$Centro de
  Radioastronom\'{i}a y Astrof\'{i}sica, UNAM Campus Morelia, Apartado
  Postal 3-72, 58090, Morelia, Michoac\'{a}n, M\'{e}xico.}

\begin{document}

\pagerange{\pageref{firstpage}--\pageref{lastpage}} \pubyear{2014}

\maketitle

\label{firstpage}

\begin{abstract}

  We carry out high resolution two-dimensional radiation-hydrodynamic
  numerical simulations to study the formation and evolution of hot
  bubbles inside planetary nebulae (PNe). We take into account the
  evolution of the stellar parameters, wind velocity and mass-loss
  rate from the final thermal pulses during the asymptotic giant
  branch (AGB) through to the post-AGB stage for a range of initial
  stellar masses. The instabilities that form at the interface between
  the hot bubble and the swept-up AGB wind shell lead to
  hydrodynamical interactions, photoevaporation flows and opacity
  variations. We explore the effects of hydrodynamical mixing combined
  with thermal conduction at this interface on the dynamics,
  photoionization, and emissivity of our models. We find that even
  models without thermal conduction mix significant amounts of mass
  into the hot bubble. When thermal conduction is not included, hot
  gas can leak through the gaps between clumps and filaments in the
  broken swept-up AGB shell and this depressurises the bubble. The
  inclusion of thermal conduction evaporates and heats material from
  the clumpy shell, which expands to seal the gaps, preventing a loss
  in bubble pressure. The dynamics of bubbles without conduction is
  dominated by the thermal pressure of the thick photoionized shell,
  while for bubbles with thermal conduction it is dominated by the
  hot, shocked wind.

\end{abstract}


\section{Introduction}
\label{sec:intro}

Planetary nebulae (PNe) are formed as the result of the evolution of
low- and intermediate-mass stars ($M_{\mathrm{ZAMS}}\sim 1$--$8
M_{\odot}$). In the most simple scenario of the formation of PNe, the
star evolves to the asymptotic giant branch (AGB) phase developing a
dusty, slow, and dense wind that interacts with the interstellar
medium \citep[ISM; e.g.,][]{Cox2012,Mauron2013}. After the star
evolves off the AGB phase towards hotter effective temperatures, the
post-AGB star develops a high ultraviolet flux that photoionizes the
AGB material. This phase is also characterized by a fast wind that is
able to sweep up, shock, and compress the photoionized AGB envelope
\citep{Kwok1978,Balick1987}. 

The first report of the detection of X-ray emission associated with a
PN was made by \citet{deKorte1985} using observations from the
\textit{EXOSAT} satellite. Since then a great number of PNe have been
observed to harbor X-ray emission. A comprehensive review of the PNe
reported with the \textit{old} X-ray facilities (\textit{Einstein},
\textit{EXOSAT}, \textit{ROSAT}, and \textit{ASCA}) was presented by
\citet{Chu2003}. Given the capabilities of those satellites, it was
not possible to unambiguously disentangle the photospheric emission
from the central star (CSPN) from that of the nebular diffuse
emission. With the unprecedented sensitivity and angular resolution of
the \textit{Chandra} and \textit{XMM-Newton} X-ray observatories this
problem can finally be resolved. By 2009, diffuse X-ray emission
delimited by bright optical bubbles had been reported for ten PNe
\citep[][and references therein]{Kastner2008}. More recently,
\citet{Kastner2012} reported that of the 35 PNe observed to that time
with \textit{Chandra}, $\sim 30$\% display diffuse X-ray emission. In
all positive detections, the X-ray emission comes from a hot bubble
inside an optical shell.

The nature of the diffuse X-ray emission from hot bubbles in PNe is
still a matter of discussion.  The main problem is that in the
wind-wind interaction scenario the shocked plasma would possess
temperatures of $T\sim10^{7}-10^{8}$~K as expected from an adiabatic
shocked wind with terminal velocities of
$v_{\infty}\gtrsim10^{3}$~km~s$^{-1}$ and plasma densities
$\lesssim$10$^{-2}$~cm$^{-3}$. Measured central star wind velocities
using P Cygni profiles of UV lines indicate that CSPN wind velocities
are generally above 1000~km~s$^{-1}$ and can be as high as
4000~km~s$^{-1}$ \citep[see][]{GuerrerodeMarco2013}. However, the
X-ray observations suggest plasma temperatures of $T_{\mathrm{X}}\sim
10^{6}$~K and electron densities of 1--10$^{2}$~cm$^{-3}$ \citep[][and
references therein]{Kastner2012}\footnote{The temperature in an
  adiabatically shocked stellar wind is defined by the terminal
  velocity as $kT=3 \mu m_{\mathrm{H}} V_{\infty}^{2} /16$, where $k$,
  $\mu$, and $m_{\mathrm{H}}$ are the Boltzmann constant, the mean
  particle mass, and hydrogen mass, respectively.}. It should be noted
that the same temperature and density discrepancy is found in hot
plasmas detected towards H\,{\sc ii} regions and Wolf-Rayet\,(WR)
bubbles \citep[e.g.,][]{Gudel2008,Toala2012}.

\citet{MellemaFrank1995} and \citet{Mellema1995} studied the hot
bubbles produced in 2D radiation-hydrodynamic simulations of the
formation and evolution of aspherical planetary nebulae. This
pioneering work showed that the interface between the hot bubble and
the swept-up shell of AGB material is unstable, but unfortunately the
numerical resolution of these simulations was not high enough to fully
develop the instabilities. These papers also estimated the soft X-ray
emission predicted by the simulations and found that it came from the
thin interface between the hot bubble and the surrounding photoionized
material where numerical diffusion spreads the temperature and density
jumps across 4 or 5 computational cells.

More recent numerical models have studied different aspects of the
formation of hot bubbles inside PNe and their associated X-ray
emission. The influence of the central star stellar wind parameters
during the post-AGB phase was studied in purely hydrodynamical,
one-dimensional models by \citet{Stute2006} and
\citet{Akashi2007}. They concluded that in this phase the fast wind
velocity must be below 10$^{3}$~km~s$^{-1}$ in order to explain such
low temperatures in the X-ray-emitting plasma, however, this is not
consistent with the results from the observed P Cygni profiles of UV
lines of central star winds. A different explanation for the diffuse
emission was put forward in an analytical approach proposed by
\citet{Lou2010}. One difficulty in these works is that they do not
take into account the time evolution of the stellar wind properties
(velocity and mass-loss rate) and the star's ionizing photon rate.

A more detailed treatment of the evolution of the stellar wind
parameters of the central star and the evolution of the X-ray emission
from the hot bubble was presented by the Potsdam group in
\citet{Steffen2008}. They started their 1D study of the evolution of
PNe with a parameter study \citep[see][]{Perinotto1998,Perinotto2004}
in which they included a large set of initial models from the AGB
phase to hot white dwarf (WD) using post-AGB stellar evolution models
from \citet{Blocker1995} and \citet{Schonberner1983}. They used three
different numerical approaches to the AGB wind evolution: a fixed
density profile $\rho(r) \propto r^{-2}$, a kinematical approach, and
a hydrodynamic calculation. In the two latter cases the AGB density
and velocity profiles depend directly on the evolutionary track
used. Specifically, they found that for the hydrodynamical approach,
the dependence of the density profile $\rho(r)$ might vary depending
on the initial mass from $r^{-1}$ to r$^{-2.5}$ with an even steeper
gradient ($r^{-3.5}$) farther from the central star. The transition
phase between the AGB and WD phase was developed using the Reimers
formulation \citep{Reimers1975}, but they concluded that the dynamical
impact of this short-lived phase is rather small. One of their
conclusions is that PNe formed from massive central stars will never
become optically thin in the high-luminosity phase unless a
low-density AGB envelope is developed.

In \citet{Steffen2008}, one-dimensional radiation-hydrodynamic models
were used to study the effect of classical and saturated thermal
conduction on the soft X-ray emission from PNe. The \textit{CHIANTI}
software was used to generate synthetic spectra and study the time
variation of the luminosities and surface brightness profiles. They
concluded that thermal conduction must be taken into account in order
to reproduce the observed X-ray luminosities in closed inner cavities
in PNe. Since thermal conduction is suppressed perpendicular to the
magnetic field direction, this implies that magnetic fields in PNe
must be absent or weak \citep[e.g.,][]{Soker1994}. These results are
consistent both with wind velocities greater than 1000~km~s$^{-1}$ and
with the observed soft X-ray emission in PNe having temperatures of a
few times $T\sim10^{6}$~K. However, the low X-ray temperatures are
only obtained if thermal conduction is included---models without
conduction fail to reproduce the observations.

An interesting scenario has been presented by \citet{Akashi2008} in
which the soft X-ray-emitting gas can be explained as the result of
the interaction of a jet with a spherical AGB wind. This model does
not take into account the evolution of the central star in any phase
nor the ionizing photon flux in the post-AGB phase. The extended soft
X-ray emission predicted by this model comes from an expanding bubble
of adiabatically cooling shocked jet material, where the jet
velocities can lie between $\sim 500$ and 3000~km~s$^{-1}$. The source
of the jet is a collimated wind from a binary companion.

This is the first of a series of papers in which we will study the
formation, characteristics and evolution of hot bubbles in PNe and
their associated soft X-ray emission. Our main goal in the present
paper is to use 2D axisymmetric numerical simulations to study the
formation of the hydrodynamical and ionization instabilities created
in the wind-wind interaction between the fast wind and the previously
ejected AGB slow wind, and the r\^ole played by these instabilities in
reducing the temperature at the edge of the diffuse hot bubble. The
importance of these instabilities has been downplayed by other authors
in earlier one-dimensional numerical treatments of this problem
\citep{Villaver2002a,Villaver2002b,Perinotto2004,Steffen2008}. We find
that the instabilities result in density inhomogeneities that become
enhanced and can form dense clumps in the region of the contact
discontinuity between the shocked fast wind and the swept-up AGB
material. Secondary shocks, photoevaporation flows and photoionization
shadows are some of the phenomena associated with the interaction of
the fast wind and the ionizing photon flux with these clumps, with
consequences both for the hot bubble and for the external photoionized
shell. We argue that this highly dynamic mixing region will be an
important source of diffuse soft X-ray emission in PNe. 
We compare the characteristics of models with different
initial masses, as more massive stars evolve faster.

\begin{figure*}
\begin{center}
\includegraphics[width=1.\linewidth]{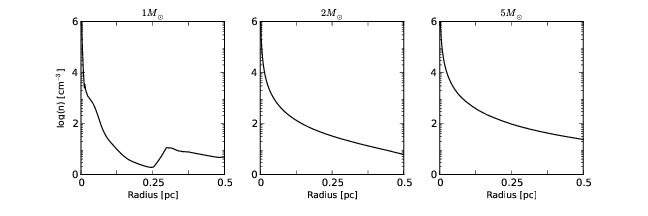}
\caption{Radial distribution of the number density at the end of the
  AGB phase for models with initial masses ($M_{\mathrm{ZAMS}}$) of 1,
  2, and 5~$M_{\odot}$ zoomed at the inner 0.5~pc as obtained from 1D
  simulations.}
\label{fig:density_1D}
\end{center} 
\end{figure*}

The structure of the paper is as follows: in \S~\ref{sec:num-method}
we describe the numerical method and the physics included in the
simulations and in \S~\ref{sec:stellar-evolution} we present the
stellar evolution histories and the stellar wind parameters and
ionizing photon rates used as input to the models. Our results are
presented in \S~\ref{sec:results} and the hot bubble properties and
evolution for the cases with and without thermal conduction are
discussed in \S~\ref{sec:discussion}. In \S~\ref{sec:summary} we
summarize and draw conclusions.

\section{Numerical Method}
\label{sec:num-method}

The radiation-hydrodynamics numerical scheme used in this work is
similar to that used in \citet{Toala2011} and \citet{Arthur2012}. The
main difference between that scheme and the one used in this paper is
the way in which the stellar wind is injected into the computational
domain.

\subsection{Overview}
First, the AGB wind is modelled in 1D spherical symmetry. This is a
slow, dense wind expanding at subsonic velocities into the surrounding
medium, which results in a $r^{-2}$ density distribution bounded by a
thin, dense shell, located at $\lesssim$2~pc as the result of the
interaction with the ISM. This shell will not be relevant for the PN
formation \citep[see figure~10 in][]{Villaver2002b}. Variations in the
AGB wind parameters during this stage due to thermal pulses can
produce internal density structure superimposed on the general
$r^{-2}$ distribution (see Fig.~\ref{fig:density_1D}). We do not
consider the proper motion of the star during this stage, which could
lead to anisotropy in the circumstellar medium
\citep[see][]{Esquivel2010,Mohamed2012, Villaver2012} prior to the
onset of the fast wind.

At the end of the AGB stage, the 1D spherically symmetric results are
remapped onto a 2D axisymmetric $(r,z)$ grid to continue the evolution
through the PN stage, where the effective temperature of the central
star increases, resulting in a two-orders of magnitude increase in the
stellar wind velocity and the onset of an ionizing photon flux.  The
combined effects of the fast wind--AGB-wind interaction, the density
gradient, and the ionizing photon flux lead to the formation of
structures at the contact discontinuity due to Rayleigh-Taylor,
thin-shell and shadowing instabilities \citep[e.g.,][]{GS1999}. These
phenomena cannot be modelled in 1D spherical symmetry. The dense
clumps formed as a result of these instabilities are ablated and
photoevaporated by their interaction with the shocked fast wind and
the ionizing photon flux. This leads to a complex scenario in which
the densities and temperatures of the material at the edge of the hot,
shocked wind bubble do not mark a sharp transition as in the 1D
case. In our models, the large surface area presented by these
structures formed at the contact discontinuity enhances the
evaporation of material from the dense, cold clumps into the hot,
tenuous bubble.

\subsection{Code Description}
In 1D spherical symmetry, the appropriate Eulerian conservation
equations are 
\begin{equation}
\frac{\partial \mathbf{U}}{\partial t} + \frac{1}{r^2}\frac{\partial
  r^2 \mathbf{F}}{\partial r} = \mathbf{S} \ ,
\end{equation}
where $\mathbf{U} = \left[\rho,\rho u,e\right]$ is the
vector of conserved quantities, $\mathbf{F} = \left[\rho u,p+\rho
  u^2,u(e+p)\right]$ is the vector of fluxes and $\mathbf{S} =
\left[0,2p/r,G-L\right]$ is the vector of source terms, both physical
and geometrical.
Here, $\rho$, $u$, $p$ are the mass density, radial velocity
and pressure, respectively, and $e$ is the total energy, where
\begin{equation}
e = \frac{p}{\gamma -1} + \frac{1}{2}\rho u^2 \ .
\end{equation}
The energy source terms $G$ and $L$ are the heating and cooling rates,
respectively, and $\gamma$ is the ratio of specific heats.  The
neutral hydrogen density is advected through the computational domain
according to the equation
\begin{equation}
\frac{\partial \rho y_0}{\partial t} +
\frac{1}{r^2}\frac{\partial}{\partial r} r^2 \rho y_0 u = 0 \ ,
\end{equation}
where $y_0$ is the H\,{\sc i} fraction.

The hydrodynamic conservation equations are solved using a
second-order, finite volume, Godunov-type scheme with outflow-only
boundary conditions at the outer boundary. At every time step, the
density, momentum and energy in an inner spherical region are reset
with a free-flowing stellar wind, whose parameters are interpolated
from a table according to the current time values derived from stellar
evolution models. The hydrogen ionizing photon rate,$S_{\star}$, is
also interpolated from these tables and the transfer of ionizing
radiation is performed by the method of short characteristics
\citep{Raga1999} in spherical symmetry. This method consists of
finding the column density of neutral hydrogen at each point in the
grid and thus the optical depth to each point from the ionizing
source. A single frequency of ionizing radiation is assumed in these
calculations.

The time-dependent hydrogen ionization balance equation takes into
account both photoionization and collisional ionization together with
recombination:
\begin{equation}
  \frac{\partial y_\mathrm{i}}{\partial t} = n_e C_\mathrm{0}
  y_\mathrm{0} + P_0 y_\mathrm{0}  - n_e R_\mathrm{i} y_\mathrm{i} \ ,
\label{eq:hionization}
\end{equation}
where $y_0$ and $y_\mathrm{i}$ are the neutral and ionized hydrogren
fractions and $n_e$ is the electron density. The collisional
ionization rate of neutral hydrogen, $C_0$, and the radiative
recombination rate of ionized hydrogen, $R_\mathrm{i}$, are functions
of temperature only and are obtained from analytic fits. The
photoionization rate in each cell, $P_0$, is calculated using the
radiation transport procedure. The ionization update is performed
after the main hydrodynamic evolution step using operator splitting.

The hydrodynamic and ionization equations are coupled through the
pressure term in the momentum and energy equations (since the gas
pressure depends on the total number of particles, i.e., ions,
neutrals and electrons), and also through the heating and cooling
source terms in the energy equation. The hydrogen ionization balance
equation (Eq.~4) is solved and other elements are assumed to be in
ionization equilibrium at the temperature and radiation conditions for
each computational cell. Although this procedure is not as accurate as
taking into account the fully time-dependent ionization of the most
important ions \citep{Steffen2008}, it has the benefit of being
computationally efficient and is easily extended to two (or more)
dimensions.

\begin{figure}
\begin{center}
\includegraphics[width=1.\linewidth]{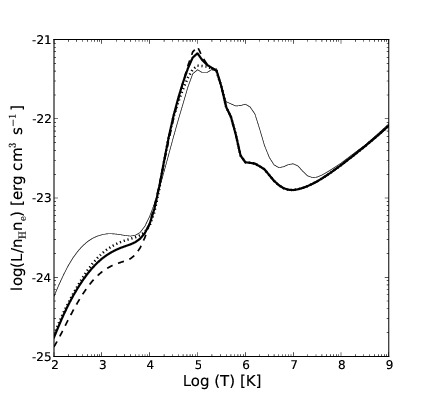}
\caption{Cooling rate per hydrogen nucleon for gas in a photoionizing
  radiation field obtained from Cloudy (see text for details). Dotted,
  solid, and dashed lines correspond to stars with effective
  temperatures of 50, 100, and 150~kK, respectively, computed with
  typical PNe abundances (see Table~1). The thin line corresponds to a
  stellar effective temperature of 40~kK and ISM abundances.}
\label{fig:cooling}
\end{center} 
\end{figure}

The radiative cooling term $L$ takes into account abundances
appropriate to PNe, via the Cloudy PN abundance set \citep[][see
Table~\ref{tab:abundances_cloudy}]{Ferland2013}. These abundances are
also used to calculate the appropriate mean nucleon mass. Cooling
rates in photoionized gas at temperatures $10^4$--$10^{4.5}$~K are
less than in collisionally ionized gas since there will be little
Ly$\alpha$ cooling due to the absence of neutral hydrogen.
We have generated a table of cooling rates using the Cloudy
photoionization code for a hot central star with an effective
temperature of 100~kK, a mean gas density of 10$^{3}$~cm$^{-3}$, and
an ionizing flux $\phi = S_{\star}/4\pi r^2 =
10^{11}$~cm$^{-2}$~s$^{-1}$ (see
Figure~\ref{fig:cooling})\footnote{The most accurate way of doing this
  would be tailoring the cooling rate to the effective temperature of
  the central star at each time and the ionizing flux at each position
  in the nebula, but this procedure would be very time
  consuming. Nevertheless, Figure~\ref{fig:cooling} shows that the
  cooling curves for stars between 50 and 150~kK are very similar.},
where $r$ is a representative value of the gas parcel radius inside
the nebula. For example, when $S_{\star} = 10^{47}$~s$^{-1}$, $r$
would be $0.1$~pc. In order to avoid spurious cooling due to numerical
diffusion at contact discontinuities, we limit cooling in cells whose
cooling is more than double that of its neighbours, as described in
\citet{GarciaArredondo2002}. The heating term, $G$, depends on the
local photoionization rate and the current stellar effective
temperature.

\begin{table}
\scriptsize
\caption{Abundance set for a PN and the interstellar medium\,(ISM) in Cloudy$^{\mathrm{a}}$}
\centering
\begin{tabular}{lcc}
\hline\hline\noalign{\smallskip}
\multicolumn{1}{l}{Atom}&
\multicolumn{1}{c}{PN} &
\multicolumn{1}{c}{ISM} \\
\multicolumn{1}{l}{}&
\multicolumn{1}{c}{(Log$X$+12)} &
\multicolumn{1}{c}{(Log$X$+12)} \\
\hline
\noalign{\smallskip}
He & 11.0  & 10.991 \\
C  & 8.892 & 8.399 \\
N  & 8.255 & 7.899\\
O  & 8.643 & 8.504\\
F  & 5.477 & 4.301 \\
Ne & 8.041 & 8.090\\
Na & 6.279 & 5.500\\
Mg & 6.204 & 7.100\\
Al & 5.431 & 4.900\\
Si & 7.00  & 6.500\\
P  & 5.301 & 5.204\\
S  & 7.00  & 7.511\\
Cl & 5.230 & 5.000\\
Ar & 6.431 & 6.450\\
K  & 5.079 & 4.041\\
Ca & 4.079 & 2.612\\
Fe & 5.699 & 5.800\\
Ni & 4.255 & 4.266\\
\hline
\end{tabular}
\begin{list}{}{}
\item{$^{\mathrm{a}}$ From \citet{Ferland2013}}
\end{list}
\label{tab:abundances_cloudy}
\end{table}

Thermal conduction is optionally included through the diffusion
equation
\begin{equation}
\rho c_\mathrm{V} \frac{\partial T_e}{\partial t} = \nabla \cdot
\left( D\nabla T_e\right) \ ,
\label{eq:diffuse}
\end{equation}
where $T_e$ is the electron thermal temperature, $c_\mathrm{V}$ is the
specific heat at constant volume and $D \propto T_e^{5/2}$ is the
diffusion coefficient. A Crank-Nicholson type scheme \citep{Press1992}
is used to solve the diffusion equation through operator splitting
after the main hydrodynamic and radiation transport steps. Saturated
conduction is treated by limiting the electron mean free path, and by
using substeps of the hydrodynamic time step, as described in
\citet{Arthur2012}.

The 1D numerical scheme is used to simulate the evolution through to
the end of the AGB stage. The AGB stage is deemed to end once the
stellar effective temperature is more than $10^4$~K and the ionizing
photon rate rises steeply. At this point, the 1D distributions of
density, momentum and total energy are remapped to a 2D cylindrically
symmetric grid using a volume-weighted averaging procedure to ensure
conservation of these quantities. The calculation then proceeds in 2D
but we also continue the 1D calculation for comparison.

In 2D cylindrical symmetry, the conservation equations in the $r-z$
plane can be written
\begin{equation}
  \frac{\partial \mathbf{U}}{\partial t} + \frac{1}{r} \frac{\partial
    r\mathbf{R}}{\partial r} + \frac{\partial \mathbf{Z}}{\partial z}
  = \mathbf{S}
\end{equation}
where, in an analogous fashion to the spherically symmetric case,
$\mathbf{U} = \left[\rho,\rho u,\rho v,e\right]$ is the vector of
conserved quantities (mass, momenta and energy), $\mathbf{R} =
\left[\rho u,p+\rho u^2,\rho uv,u(e+p)\right]$ and $\mathbf{Z} =
\left[\rho v,\rho uv, p+\rho v^2,v(e+p)\right]$ are the flux vectors,
and $\mathbf{S} = \left[0,p/r,0,G-L\right]$ is the vector of
geometrical and physical source terms. Here, $u$ is the radial
velocity, $v$ is the $z$-velocity, and the total energy $e$ is now
defined by
\begin{equation}
  e = \frac{p}{\gamma - 1} + \frac{1}{2}\rho (u^2 +
  v^2) \ .
\end{equation}
The $r$ and $z$ directions are dealt with separately using operator
splitting.

The 2D cylindrically symmetric hydrodynamic conservation equations,
are solved using a hybrid scheme, in which alternate time steps are
calculated by a second-order Godunov-type method and the van Leer
flux-vector splitting method \citep{vanLeer1982}. The outer $r$ and
$z$ boundaries are outflow only, while the radial fluxes are zero at the
symmetry axis.

\begin{figure}
\includegraphics[width=1.0\linewidth]{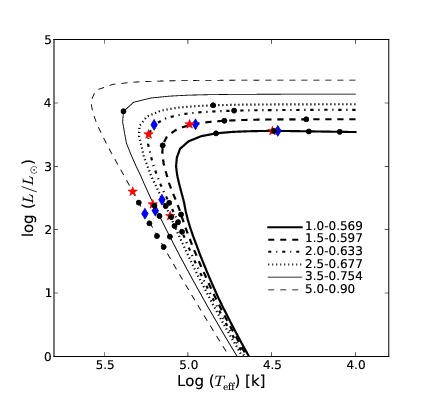}
\caption{Evolutionary tracks for the WD models from
  \citet{Vassiliadis1994} used in this paper. The positions of the
  (black) circles mark times at [1, 5, 10, and 20]$\times10^{3}$~yr on
  the tracks. Stars (red) and diamonds (blue) indicate the positions
  on the tracks when the 2D hot bubble has a mean radius of
  $\sim0.2$~pc (or the maximum extension in the case of the most
  massive models), for cases without and with thermal conduction,
  respectively.}
\label{fig:VW_tracks}
\end{figure} 

In an analogous fashion to the 1D case, the stellar wind is input into
the computational domain as a free-flowing wind in an inner spherical
region, which is reset at every time step with the appropriate values
interpolated from the stellar evolution table. The radiation transport
is performed by the 2D axisymmetric version of the same short
characteristics method and the source of ionization (the star) is
taken to be positioned on the symmetry axis. In the same way, the
hydrodynamic and ionization equations are coupled through the pressure
term in the momenta and energy equations and through the source term
in the energy equation. We again solve the time-dependent hydrogen
ionization equation (Eq.~\ref{eq:hionization}) and employ the 2D
advection equation to describe the transport of the neutral hydrogen
fraction through the computational domain. We use the same radiative
cooling source term and again limit the cooling in a cell whose
cooling is more than double that of its neighbours in an effort to
avoid spurious cooling due to numerical diffusion in the vicinity of
contact discontinuities. We optionally include thermal conduction
through the axisymmetric diffusion equation (Eq.~\ref{eq:diffuse}),
which is again solved by appropriate numerical techniques through
operator splitting after the main hydrodynamic and radiation transport
steps.

\subsection{Computational grid and initial conditions}
For the 1D spherically symmetric calculations, the initial conditions
for the interstellar medium are a density of $n_0 = 1$~cm$^{-3}$ and a
temperature of $T_0 = 100$~K. The initial grid has 2000 uniformly
spaced radial cells and the initial spatial size is 0.5~pc. The wind
injection zone comprises the innermost 10 cells.  The computer code
automatically detects when the outer edge of the expanding AGB wind
has become close to the outer boundary and incrementally extends the
computational grid with more cells at the original initial conditions
of density and temperature whenever necessary. This procedure avoids
large computational domains at the beginning of the simulation and
also means that the final size depends on the particular stellar
evolution model being calculated without needing to compromise the
initial resolution.

The 2D axisymmetric calculations are performed on a fixed grid of 1000
radial by 2000 $z$-direction cells of uniform cell size and total grid
spatial extent $0.5 \times 1$~pc$^2$. The initial conditions for these
calculations are the remapped results of the innermost 0.5~pc of the
1D calculations, which entirely fill the 2D computational domain.  The
free-wind injection region has a radius of 40 cells, corresponding to
0.02~pc. With this number of cells, the spherical free-wind injection
region can be adequately represented on the axisymmetric grid. The use
of a fixed-size grid has the disadvantage that, for the lowest initial
stellar mass models, the outer photoionized shell leaves the
computational domain at late times. However, since we are primarily
interested in the formation of hot bubbles and their potential for
emitting soft X-rays, we have decided to focus on the region $\left(
  r^2 + z^2 \right)^{0.5} < 0.5$~pc.

The remap to the 2D grid also downgrades the numerical resolution
since the 2D calculations are computationally far more expensive than
the 1D calculations. In order to provide a meaningful comparison with
1D results, we also downgrade the resolution of the continuing 1D
simulations and fix the total number of radial cells to 1000.

\section{Stellar Evolution Models}
\label{sec:stellar-evolution}

\begin{figure*}
\includegraphics[width=.5\linewidth]{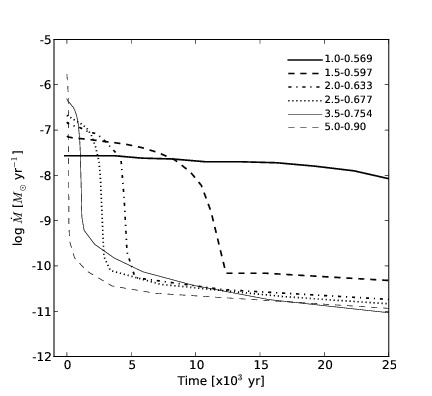}~
\includegraphics[width=.5\linewidth]{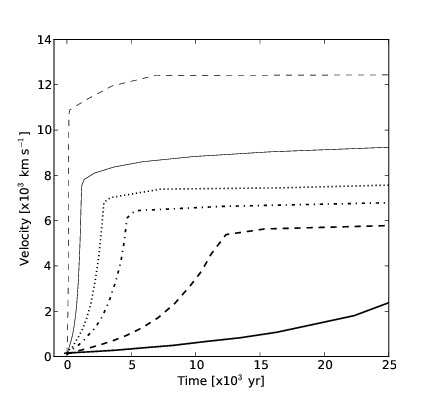}\\
\includegraphics[width=.5\linewidth]{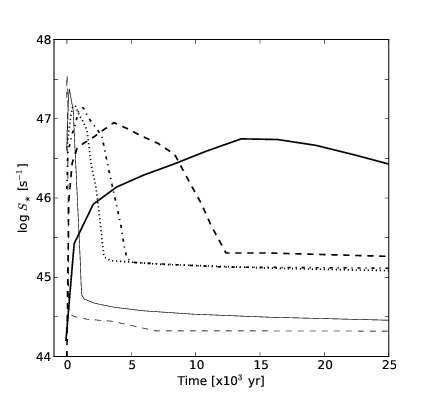}~
\includegraphics[width=.5\linewidth]{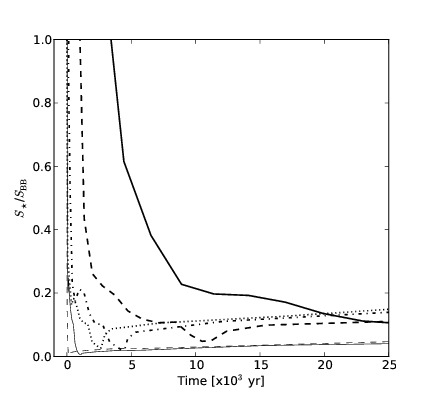}\\
\caption{(Top-left) Mass-loss rate, (top-right) stellar wind velocity,
  and (bottom-left) ionizing photon flux for the different stellar
  models. (Bottom-right) Ratio of the ionizing photon flux over that
  expected from a black body model. The initial-final mass of each
  model is marked on top-left panel.}
  \label{fig:wind_parameters}
\end{figure*}

In this paper we use publicly available stellar evolution models with
initial masses of 1, 1.5, 2, 2.5, 3.5, and 5~$M_{\odot}$ at solar
metallicity ($Z=0.016$) from \citet{Vassiliadis1993} for the AGB phase
and from \citet{Vassiliadis1994} for the post-AGB phase (see
Figure~\ref{fig:VW_tracks}). These models correspond to final WD
masses of 0.569, 0.597, 0.633, 0.677, 0.754, and 0.90~$M_{\odot}$. We
will label each set of simulation as a combination of their initial
masses and the final WD mass, for example, the model with initial mass
of 1~$M_{\odot}$ and final WD mass of 0.569~$M_{\odot}$ will be called
1.0-0.569. Thus, the full set of models will be referred as 1.0-0.569,
1.5-0.597, 2.0-0.633, 2.5-0.677, 3.5-0.754, and 5.0-0.90.

The AGB wind parameters (velocity and mass-loss rate) are empirical
relations derived by \citet{Vassiliadis1993} from observations of
Galactic and Large Magellanic Cloud Mira variables and dust-enshrouded
AGB stars and are the same as those used by \citet{Villaver2002a}. The
stellar wind parameters and ionizing photon flux during the hot CSPN
stage were computed using the WM-basic hot star stellar atmosphere
code
\citep{Pauldrach1986,Pauldrach1987,Pauldrach1994,Pauldrach2001,Pauldrach2012}
using the stellar effective temperature, radius and surface gravity
taken at discrete times from the stellar evolution models from
\citet{Vassiliadis1994}. Figure~\ref{fig:wind_parameters} (top panels)
shows the mass-loss rates and velocity profiles corresponding to the
CSPN stage for the six evolution models. An improvement between our
stellar parameters and those from previous numerical studies is that
we do not model the ionizing photon flux as black body radiation. With
the use of WM-basic we obtained realistic spectra at different
evolution times of the star and we integrated them to compute the
ionizing photon rate. Figure~\ref{fig:wind_parameters} (bottom-left
panel) shows the resultant integrated ionizing photon flux for the six
models studied here. As the star becomes hotter and more compact, the
opacity increases and the ionizing photon flux diminishes considerably
compared to a black body (see Figure~\ref{fig:wind_parameters}
bottom-right panel).

As can be seen from Figure~\ref{fig:wind_parameters}, the potential
lifetime of a PN produced by a star of initial mass 5.0-0.90~$M_\odot$
is considerably shorter than that of a 1.0-0.569~$M_\odot$ star.

In common with \citet{Villaver2002b}, we assume that the post-AGB
stage starts when the star's effective temperature has reached
$10^4$~K and we do not model the short transition phase between the
superwind and the post-AGB phase. We are aware that the inclusion of
this transition may affect the hydrodynamics at the edge of the hot
bubble \citep{Perinotto2004}.

\section{Results}
\label{sec:results}

We will separate the presentation of the results into two groups
according to similarities in the morphological evolution. This is
mainly a consequence of the density structure at the end of the AGB
stage (see Fig.~\ref{fig:density_1D}) and the duration of the CSPN
stage. Group A comprises the results from 1.0-0.569$M_\odot$,
1.5-0.597~$M_\odot$, 2.0-0.633~$M_\odot$, and 2.5-0.677~$M_\odot$
stars, while Group B consists of the 3.5-0.754~$M_\odot$ and
5.0-0.90~$M_\odot$ stars. The Group A models start with a
circumstellar medium whose density falls off quite steeply with
distance from the star. Group B models have a less steep fall off in
the initial density distribution and are therefore denser at a given
radius. At early times, all the models create well-defined hot bubbles
with sharp shells but soon differences in the stellar wind parameters,
the circumstellar medium left by the AGB stage, the ionizing photon
rate variation and the different evolution timescales lead to
significant morphological differences. We first describe the results
without thermal conduction, and then discuss the main features of the
inclusion of this physical effect.

\begin{figure*}
\includegraphics[width=1.0\linewidth]{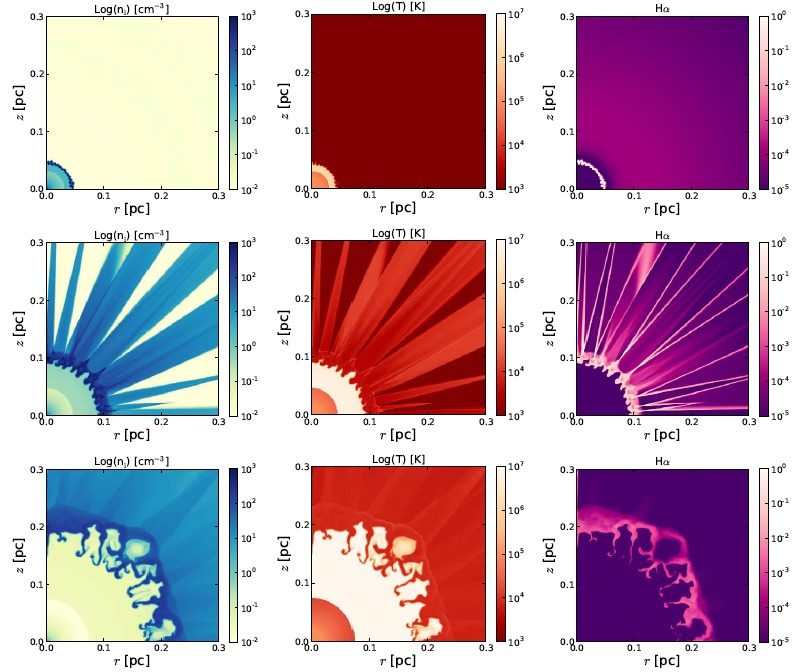}
\caption{Total ionized number density (left panels),
  temperature(middle panels) and normalized H$\alpha$ volume
  emissivity (right panels) for the PN formation for the 1.5-0.597
  model 2D simulations without thermal conduction. The rows correspond
  to the hot bubble having a mean radius of, from top to bottom, 0.05,
  0.1 and 0.2~pc. These are equivalent to 1000, 3500, and 7400~yrs of
  post-AGB evolution for this CSPN. The bottom row corresponds to the
  time when the central star has an effective temperature and
  luminosity of $T_{\mathrm{eff}}=97720$~K and $L=4310 L_{\odot}$.}
\label{fig:PN15_2D}
\end{figure*}

\begin{figure*}
\includegraphics[width=1.0\linewidth]{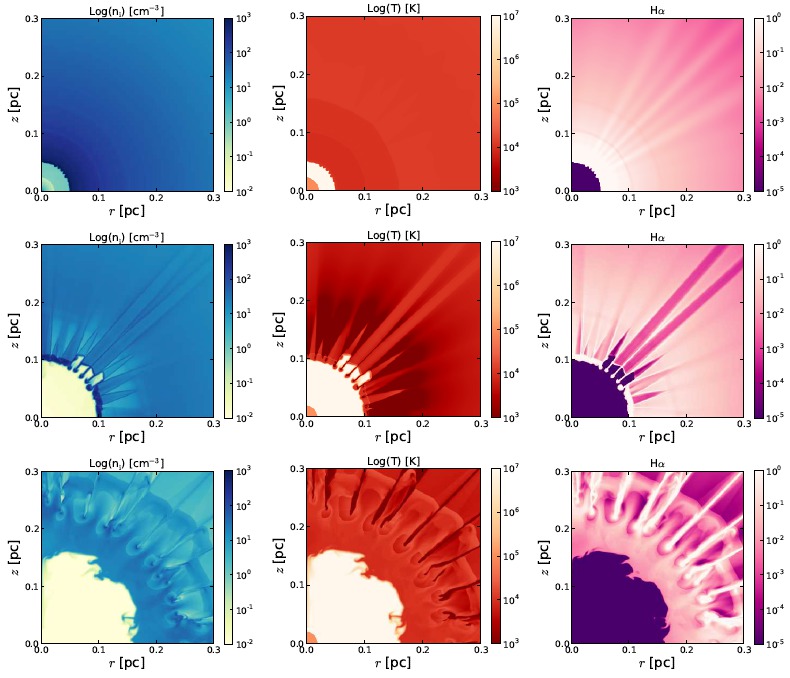}
\caption{Same as Figure~\ref{fig:PN15_2D} but for the case of
  2.5-0.677 model without thermal conduction. The rows correspond to
  the hot bubble having a mean radius of, from top to bottom, 0.05,
  0.1 and 0.2~pc. These are equivalent to 1800, 2930, and 7300~yrs of
  post-AGB evolution for this CSPN. The bottom row corresponds to the
  time when the central star has an effective temperature and
  luminosity of $T_{\mathrm{eff}}=127640$~K and $L=165 L_{\odot}$.}
\label{fig:PN25_2D}
\end{figure*}

\begin{figure*}
\includegraphics[width=1.0\linewidth]{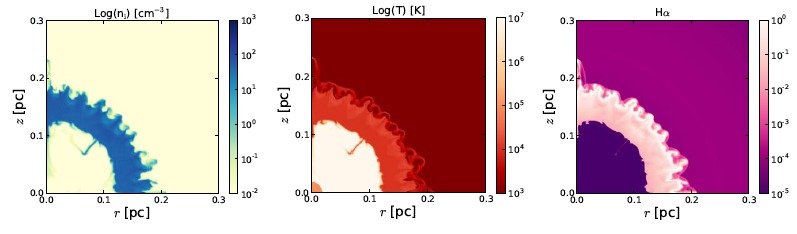}
\caption{Same as Figure~\ref{fig:PN15_2D} but for the case of
  3.5-0.754 model without thermal conduction. This set of panels
  corresponds to the maximum radius of the hot bubble, which occurs
  after 4900~yrs of post-AGB evolution for this CSPN. The panels
  corresponds to the time when the central star has an effective
  temperature and luminosity of $T_{\mathrm{eff}}=162180$~K and $L=250
  L_{\odot}$.}
\label{fig:PN35_2D}
\end{figure*}

All our models focus on the innermost 0.5~pc of the PN, since this is
the region that hosts the hot bubble. As a result, we do not follow
the complete evolution of the PN, since at late times the photoionized
outer shell leaves the computational domain.  We present a series of
images for each model, corresponding to the inner hot shocked wind
bubble having mean radius 0.05, 0.1 and 0.2~pc for Group A models, and
for the Group B models we show results corresponding to the maximum
radius of the hot, shocked bubble, since in these cases the hot bubble
stalls before it reaches 0.2~pc and then begins to collapse. In order
to highlight the main features of each group, we show figures for
representative models in the main text, and figures for the other
models are collected in Appendix~\ref{app:appa}. For Group A, the
representatives are 1.5-0.597~$M_\odot$ and 2.5-0.677~$M_\odot$ (see
Figs.~\ref{fig:PN15_2D} and \ref{fig:PN25_2D}), while for Group B the
representative is 3.5-0.754~$M_\odot$ (see Fig.~\ref{fig:PN35_2D}).

In each set of images, the total ionized number density ($n_\mathrm{i}
= \rho y_\mathrm{i}/\mu m_\mathrm{H}$, where $\mu$ mean particle
mass), temperature, and H$\alpha$ emitting region are
shown\footnote{All 2D images were produced with the matplotlib Python
  routines developed by \citet{Hunter2007}}. Although the calculations
are performed on the full $r-z$ plane, the results are symmetric about
the midplane and for reasons of space, we show only half of the
computational domain.

%%%%%%%%%%%%%%%%%%%%%%%%%%
%% AQUI LOS CON CONDUCCION

\begin{figure*}
\includegraphics[width=1.0\linewidth]{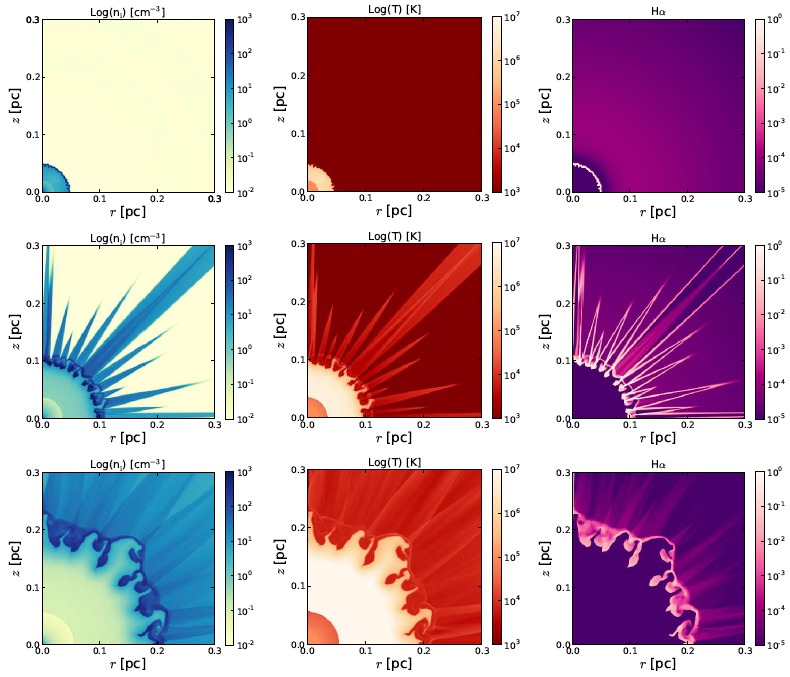}
\caption{Same as Figure~\ref{fig:PN15_2D} but for the 1.5-0.597 model
  with thermal conduction. The rows correspond to the hot bubble
  having a mean radius of, from top to bottom, 0.05, 0.1 and
  0.2~pc. These are equivalent to 1000, 3200, and 6200~yrs of post-AGB
  evolution for this CSPN. The bottom row corresponds to the time when
  the central star has an effective temperature and luminosity of
  $T_{\mathrm{eff}}=90160$~K and $L=4610L_{\odot}$.}
\label{fig:PN15_2D_cond}
\end{figure*}

\begin{figure*}
\includegraphics[width=1.0\linewidth]{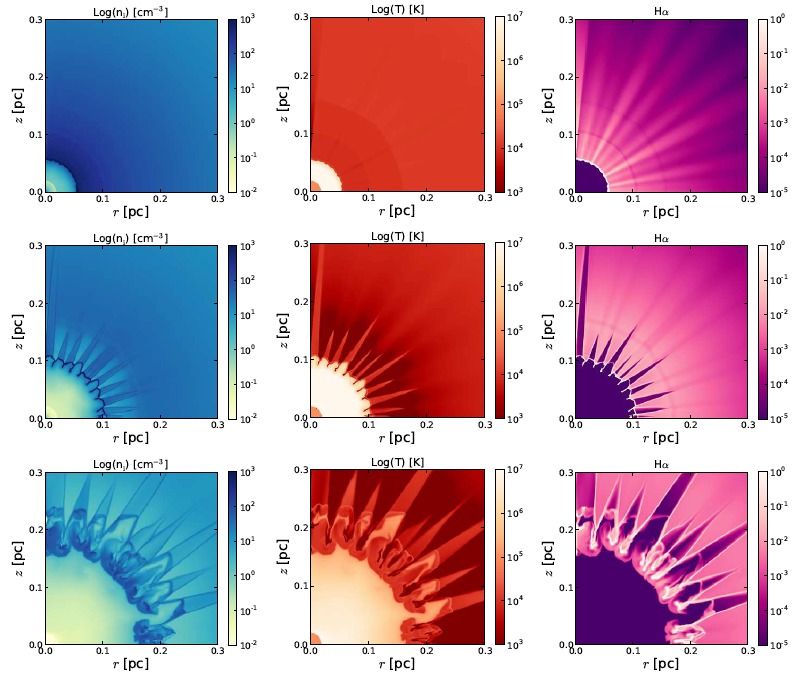}
\caption{Same as Figure~\ref{fig:PN15_2D} but for the case of
  2.5-0.677 model with thermal conduction. The rows correspond to the
  hot bubble having a mean radius of, from top to bottom, 0.05, 0.1
  and 0.2~pc. These are equivalent to 1800, 2500, and 4300~yrs of
  post-AGB evolution for this CSPN. The bottom row corresponds to the
  time when the central star has an effective temperature and
  luminosity of $T_{\mathrm{eff}}=142900$~K and $L=295 L_{\odot}$.}
\label{fig:PN25_2D_cond}
\end{figure*}

\begin{figure*}
\includegraphics[width=1.0\linewidth]{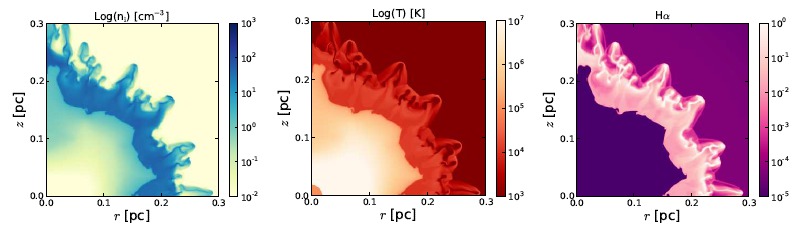}
\caption{Same as Figure~\ref{fig:PN15_2D} but for the case of
  3.5-0.754 model with thermal conduction. This set of panels
  corresponds to the maximum radius of the hot bubble, which occurs
  after 5500~yrs of post-AGB evolution for this CSPN. The panels
  corresponds to the time when the central star has an effective
  temperature and luminosity of $T_{\mathrm{eff}}=158500$~K and $L=240
  L_{\odot}$.}
\label{fig:PN35_2D_cond}
\end{figure*}

\subsection{Group A}
Models 1.0-0.569, 1.5-0.597, 2.0-0.633, and 2.5-0.677 evolve in very
similar ways due to the similarity of the initial circumstellar medium
density distribution and velocity at the start of the post-AGB phase
(see Fig.~\ref{fig:density_1D}). These models follow the classic PNe
formation scheme: the fast CSPN wind expands into the slowly expanding
$v_\mathrm{exp} \lesssim 15$~km~s$^{-1}$, smooth $\rho \propto r^{-2}$
(or steeper) density profile left after the AGB stage. The interaction
between the fast wind and the slow wind forms a two-shock pattern: the
outer shock sweeps up and compresses the AGB material, while the inner
shock brakes and heats the fast wind. The two shocks are separated by
a contact discontinuity. Behind the outer shock, the density is very
high and cooling is efficient. The thin-shell instability
\citep{Vishniac1983,Vishniac1989} acts at early times to corrugate the
contact discontinuity separating the cold, dense material from the hot
bubble and clumps begin to form (see the upper panels in
Figs.~\ref{fig:PN15_2D} and \ref{fig:PN25_2D}).

The ionization front is initially trapped close to the thin shell, but
as the shell moves outwards the opacity falls and the ionization front
progress outwards. The clumps formed by the instability lead to
variations in the opacity and as a consequence the shadowing
instability becomes important at later times
\citep{Williams1999}. Characteristic rays of alternating ionized and
neutral material form at intermediate times. These can be visualized
as bright streaks in the H$\alpha$ volume emissivity trailing from
bright H$\alpha$ clumps, as can be seen in the central panels of
Figs.~\ref{fig:PN15_2D}, \ref{fig:PN25_2D} and the Group A models in
Appendix~\ref{app:appa}. At later times, the opacity in the ring of
clumps drops as they expand outwards, and the whole outer region
becomes photoionized and more homogeneous.

The continued acceleration of the fast stellar wind means that the
swept-up shell is accelerated down the density gradient and
Rayleigh-Taylor instabilities become important at the contact
discontinuity between the low density, shocked fast wind material and
the dense, swept-up shell. This leads to the formation of elongated,
dense, partially ionized structures pointing inwards into the hot
bubble. Neutral, dense material inside these filaments can be
photoevaporated by the ionizing flux. The expanding, photoevaporated
gas flows away from the head of the neutral filament, towards the
central star. It then interacts dynamically with the radially
outflowing, shocked fast wind, giving rise to pockets of $\sim 10^6$~K
gas with densities in the range $0.1 < n_\mathrm{i} <
1.0$~cm$^{-3}$. The hot bubble, on the other hand, has temperature $>
10^7$~K and density $n_\mathrm{i} < 0.01$~cm$^{-3}$. Even when there
is no neutral material inside the dense filaments to generate
photoevaporated flows, the diversion of the shocked fast wind material
around the dense obstacles leads to hydrodynamic ablation and the
mixing of cooler, dense material into the faster, hot flow (see, e.g.,
\citealp{Hartquist1986}). In general, the models that evolve more
slowly generate longer filamentary structures, simply because they
have had a much longer time to grow by the time the hot bubble has
expanded to 0.2~pc radius.

The 2.5-0.677 case is particularly interesting. The ring of clumps
forms due to the thin-shell instability as for the other cases in this
group. However, the faster acceleration of the stellar wind and the
resulting higher pressure in the hot bubble mean that the clumps are
pushed out to about 0.1~pc before the photoionization has the
opportunity to cause this shell of swept-up material to expand. The
clumps are therefore denser in this model at this radius than for
other models. The ionizing photon rate of the central star now drops
sharply and so the dense clump material recombines. This leads to the
situation where the heads of the now neutral clumps develop ionization
fronts and photoevaporation flows pointing towards the central
star. Initially, the pressure of the hot bubble is higher than the
pressure in the photoionized material, and the photoevaporated
material is pushed back into the shell.  As the whole structure
expands slowly outwards, however, and the mass-loss rate of the
central star drops sharply, the pressure in the photoevaporated flows
becomes greater than that in the hot bubble and starts to expand away
from the neutral clumps, back towards the central star after passing
through a termination shock. This leads to the situation in the final
row of Figure~\ref{fig:PN25_2D}, where the hot bubble is surrounded by
a thick shell of photoionized material, within which are embedded
neutral clumps. The thick shell is slowly moving outwards, broadening
as it goes, and the hot bubble is confined by the slowly decreasing
pressure of the photoionized gas.

Structures similar to those shown in Figures~\ref{fig:PN15_2D},
\ref{fig:PN25_2D}, and \ref{fig:PNall_2D}, i.e. a hot
inner cavity surrounded by lacy structures in turn surrounded by a
smooth photoionized shell, are observed in \textit{Hubble Space
  Telescope} (HST) images of, e.g., IC\,418, NGC\,2392, NGC\,6826, and
NGC\,7662, which also display inner diffuse X-ray emission surrounded
by a photoionized shell detected with \textit{Chandra}
\citep{Kastner2012,Ruiz2013}.

\subsection{Group~B (3.5-0.754 and 5.0-0.90)}

Models from Group~B are the result of the evolution of the most
massive stars of our sample. The density distribution around these
stars at the end of the AGB stage does not have such a steep radial
dependence. This means that the fast wind expands into a denser medium
and the swept-up shell of AGB material is correspondingly denser. In
fact, the ionization front becomes trapped in the shell and does not
manage to break out and hence the surrounding medium remains neutral
throughout the evolution. The photoionized shell has a high thermal
pressure and effectively confines the hot bubble to small radii. This
effect is enhanced by the sharp CSPN mass-loss rate fall off (see
Fig.~\ref{fig:wind_parameters} top-left panel) and as a consequence,
the thermal pressure of the hot bubble decreases with time. There
comes a point when the pressure inside the hot bubble becomes lower
than that in the surrounding photoionized shell. This marks the point
at which the hot bubble ceases to expand and starts to collapse. In
Figures~\ref{fig:PN35_2D} and \ref{fig:PNall_2D} we show the maximum
extent of the hot bubble for these models. As can be seen, the hot
bubble is surrounded by a thick photoionized shell and this is
embedded within a dense neutral envelope.

\subsection{Thermal Conduction Results}

Figures~\ref{fig:PN15_2D_cond}, \ref{fig:PN25_2D_cond}, and
\ref{fig:PN35_2D_cond} (and see also Fig.~\ref{fig:PNall_2D_cond},
Appendix~\ref{app:appa}) show the corresponding results for the cases
with thermal conduction. Conduction by thermal electrons heats
material from the dense, cool, swept-up AGB shell and it then expands
into the hot bubble. The evaporation process results in a sheath of
comparatively dense ($n_\mathrm{i} \sim 1$~cm$^{-3}$), but hot ($10^5
< T < 10^6$~K) material around the fingers and clumps that were formed
by the different instabilities at the interface between the hot bubble
and dense swept-up shell. The total surface area of the conducting
layer is very large because of these instabilities and substantial
amounts of AGB shell material are evaporated into the hot bubble.

We find that the pressure in the hot bubble is higher in the models
with conduction and so the expansion for all models is more rapid than
for the models without conduction. This is because the evaporated
material expands into the interclump region between the clumps and
filaments in the broken shell and the bubble does not depressurize. At
a given time, the instabilities are much more developed for the models
with conduction than those without. The swept-up photoionized shells
in the models with conduction are thinner because the hot bubble
dominates the dynamics. The higher pressure means that these PNe are
essentially wind-blown bubbles surrounded by thin swept-up shells. The
models without conduction, on the other hand, have their dynamics
dictated by the pressure in the broad photoionized shell \citep[see,
  e.g.,][]{Arthur2012,Raga2012}.

\subsection{Comparison between 2D and 1D results}
\begin{figure*}
\includegraphics[width=0.33\linewidth]{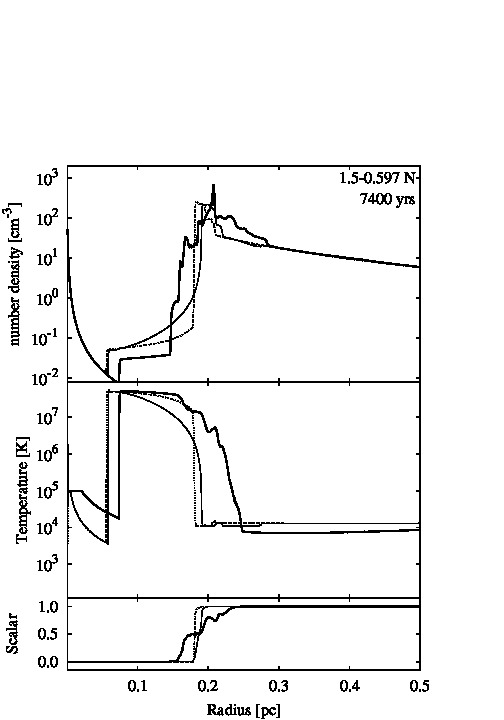}~%
\includegraphics[width=0.33\linewidth]{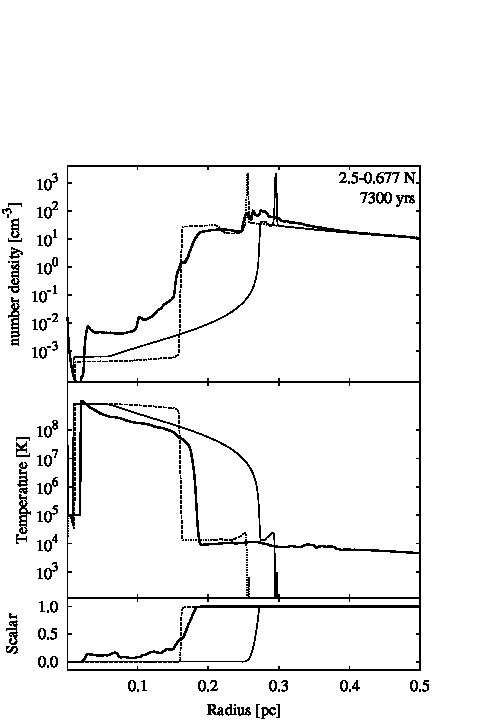}~%
\includegraphics[width=0.33\linewidth]{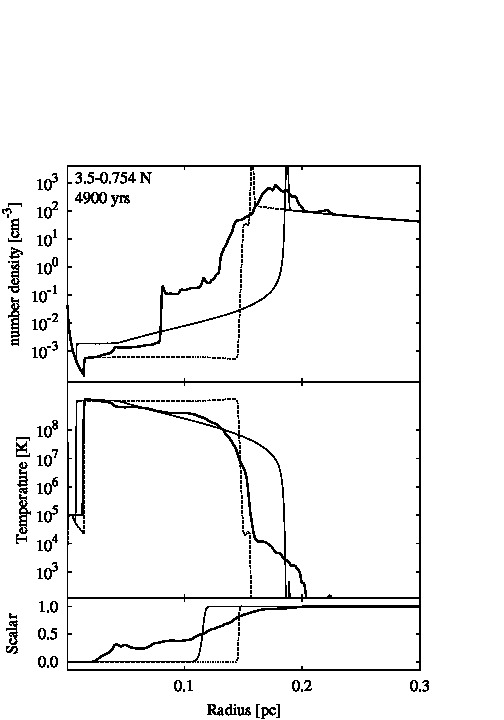}\\
\includegraphics[width=0.33\linewidth]{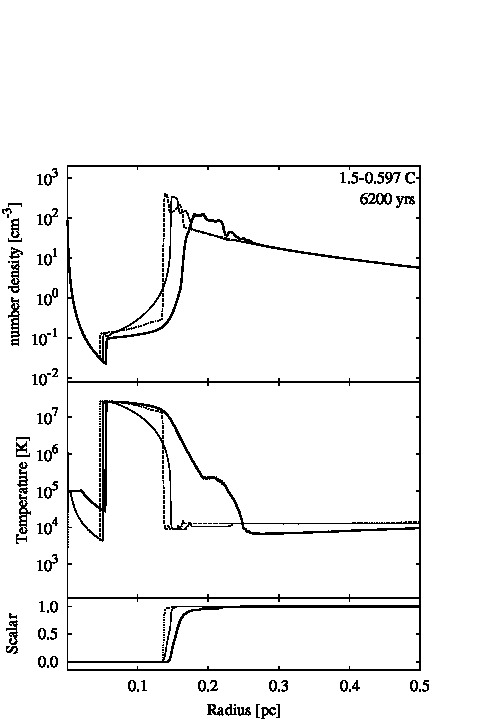}~%
\includegraphics[width=0.33\linewidth]{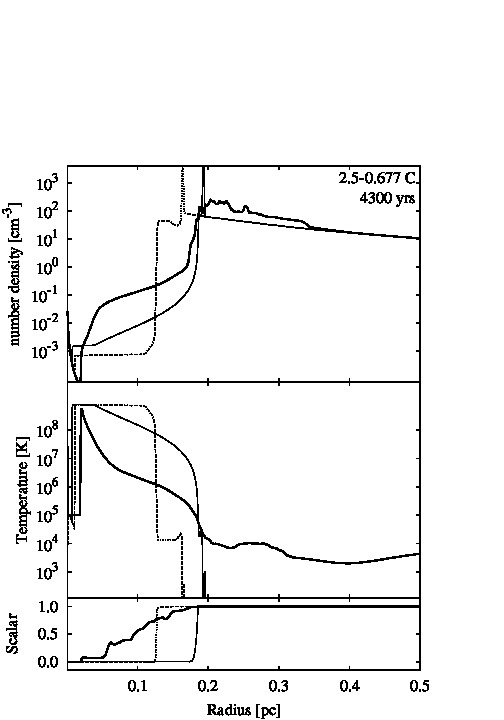}~%
\includegraphics[width=0.33\linewidth]{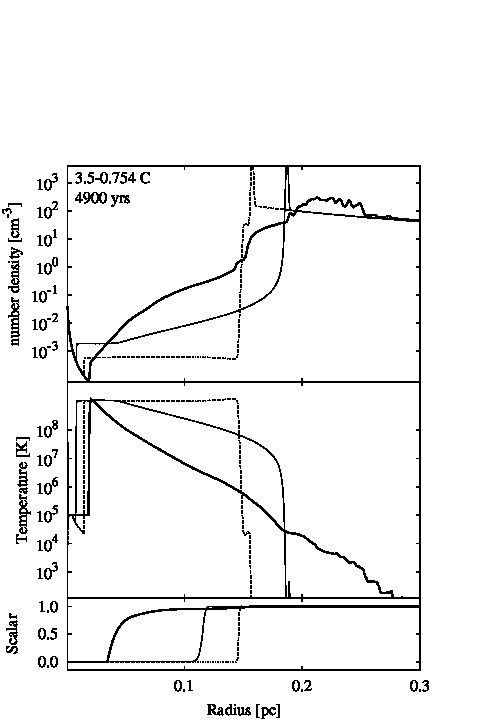}
\caption{Comparison of 2D results with 1D results. Top row---2D
  results without thermal conduction. Bottom row---2D results with
  thermal conduction. In each plot the upper panel shows the total
  number density ($n =\rho/\mu m_\mathrm{H}$) and the centre panel
  shows the temperature. The lower panel shows the value of a passive
  advected scalar. The thick solid line is the 2D simulation as a
  function of radius averaged over all angles from the stellar
  position. The thin solid line is the 1D result with thermal
  conduction and the dotted line is the 1D result without thermal
  conduction, both at the same evolution time as the corresponding 2D
  result.}
\label{fig:compare1d2d}
\end{figure*}

In Figure~\ref{fig:compare1d2d} we compare the 2D results shown in the
previous sections to the corresponding 1D results calculated at the
same numerical resolution and with the same physical processes. In
order to make the comparison, we plot the 2D results as a function of
radial distance from the star averaged over all angles. The top row of
figures shows the averaged 2D results without conduction for the three
principal models: 1.5-0.597, 2.5-0.677 and 3.5-0.754 and the bottom
row shows the results when thermal conduction is included. Each figure
also shows the 1D simulation results both with and without conduction
at the same evolution times. The upper panel in each figure is the
total number density ($n =\rho/\mu m_\mathrm{H}$), the centre panel is
the temperature, and the lower panel shows the value of a passive
advected scalar. This scalar is assigned a value of one in the AGB
wind material prior to the onset of the fast wind, and is zero in the
post-AGB fast-wind material that enters the grid
thereafter. Intermediate values indicate mixed regions: for the 1D
simulations this is an indication of a cooling region, for the 2D
simulations it is a result of the averaging process. The scalar shows
how far the evaporated dense shell material has penetrated into the
hot bubble.

We begin with an explanation of the 1D results. The classical
description of a stellar wind bubble divides it into distinct regions
(see e.g., \citealp{DysonWilliams}, \citealp{Arthur2012}). Closest to
the star, there is an inner region of unshocked, free-flowing stellar
wind, which is separated from the hot bubble by an inward-facing
shock. The hot bubble comprises hot, shocked stellar wind material and
is separated by a contact discontinuity from the swept-up shell. This
shell is composed of swept-up ambient medium, which for our models is
AGB material. There can be a photoionized region inside the swept-up
shell, in which case there will be an ionization front preceded by a
shock in the neutral medium within the shell. External to the shell is
undisturbed ambient medium. For a steady stellar wind, the pressure
between the inner stellar wind shock and the outer neutral shock will
be uniform. The contact discontinuity between the hot, shocked stellar
wind and the swept-up, photoionized shell separates diffuse gas at
$\sim 10^7$~K from dense gas at $\sim 10^4$~K. If magnetic fields are
not important at this interface, conduction due to thermal electrons
from the hot gas can lead to diffusion of heat across the contact
discontinuity (see Eq.~\ref{eq:diffuse}). For the hot, diffuse gas,
the heat diffusion leads to a loss of thermal energy and hence a
reduction of temperature in this gas together with a corresponding
increase in density, since the pressure remains constant. For the
cooler, dense gas, there is an increase in temperature. Depending on
the details of the radiative cooling, the stellar wind, and the
ionizing photon rate, one of the following must occur:
\begin{enumerate}
\item The cooling timescale for the heated dense material is so short
  that it cools immediately back to its original state. We remark that
  this case is also likely to occur if there is over-cooling at the
  contact discontinuity as a result of numerical diffusion.
\item The heated dense gas expands at constant pressure, where the
  pressure is dictated by conditions in the hot shocked wind
  bubble. The increase in volume (and corresponding decrease in
  density) in this gas pushes the shell outwards.
\item The heated dense gas cannot expand outwards and instead pushes
  inwards, increasing the pressure inside the hot, shocked bubble and
  pushing the inner stellar-wind shock closer to the star. This occurs
  particularly if the stellar wind mechanical energy is decreasing
  rapidly with time.
\end{enumerate}

Whichever of these 3 behaviours prevails initially, radiative cooling
will play an important r\^{o}le on the timescale of a cooling
time. Cooling times are shortest in the heated dense region. As this
gas begins to cool, its volume decreases (since the pressure must
remain constant because this gas is located between the hot, shocked
bubble and the surrounding photoionized region). Due to the short
evolution timescales of these planetary nebula models, cooling is not
necessarily complete on the timescale of the simulations. We see
evidence for a cooling region in the 1.5-0.597 1D model results (see
Fig.~11). Cooling in the reduced temperature diffuse gas occurs on
much longer timescales \citep[see e.g.,][]{Arthur2012}.

Figure~\ref{fig:compare1d2d} shows examples of the second and third
types of situations listed above. For example, the 1.5-0.597 models
are examples of the second case, where in the models with conduction,
the heated, dense gas expands and pushes the dense shell outwards, in
comparison with the models without conduction.  In these models, the
ambient medium gas is photoionized beyond the swept-up shell. The
3.5-0.754 models, on the other hand, are examples of the third case,
where the heated dense gas expands quite a long way back towards the
star, pushing the inner stellar wind shock to smaller radii in the
models with conduction, in comparison with the models without
conduction. In these models, the swept-up shell of ambient medium gas
is mainly neutral, since the ionizing photon rate falls off rapidly
within the first 1000~yrs of the planetary nebula evolution. The
scalar in these figures indicates the position of the interface
between the wind material and the ambient material. The 2.5-0.677
models are more complicated, since the swept-up shell contains the
ionization front and preceding neutral shock and is mainly, though not
completely, photoionized. The photoionized gas regulates the pressure
throughout the planetary nebula once the stellar wind mechanical
luminosity begins to decline.

We now examine the differences between the 2D and 1D
simulations\footnote{The obvious difference at small radii is due to
  the difference in size of the wind injection region in the 2D and 1D
  cases.}. For the 1.5-0.597 case without conduction we immediately
notice that the inner stellar wind shock for the angle-averaged 2D
results is at a larger radius than for the equivalent 1D models. This
means that the density, and hence the pressure in the hot, shocked
wind bubble is lower in the 2D case, since the post-shock temperature
is the same in all cases because it depends only on the stellar wind
velocity. The reason for the lower pressure can be appreciated by
looking at Figure~\ref{fig:PN15_2D}. The swept-up shell in this figure
has broken up due to instabilities and the hot wind can find its way
through the gaps, depressurising the hot bubble. Although the peak
density occurs at roughly the same radius as in the 1D models, warm
($T > 10^4$~K) gas extends beyond the shell as a result of mass
pick-up from the clumps and filaments by the hot wind as it flows
through the gaps in the leaky bubble. For the 1.5-0.597 2D model with
conduction, the inner stellar wind shock coincides more-or-less with
the 1D results. This is because conduction heats the gas at the
surface of the cold, dense filaments and the expansion of this gas
seals the gaps in the bubble. The result is a broad, roughly uniform
density shell with a temperature of a few times $10^5$~K.

The averaged locations of the interface between hot bubble and
swept-up shell in 2.5-0.677 2D models with and without conduction
correspond roughly to their 1D counterparts. However, the average
density as a function of radius within the hot bubble is an order of
magnitude higher in both cases. This is a result of hydrodynamical
mixing of dense material into the hot diffuse gas at the irregular
interface. Non-radial motions thoroughly mix the ablated, cool, dense
gas throughout the hot bubble, thereby raising the average density and
lowering the average temperature. This is in contrast to the 1D
conduction results, where the temperature in the hot bubble is reduced
solely as a result of heat loss due to diffusion of heat, with no
corresponding diffusion of material. The averaged 2D models have lower
density, broader, swept-up shells than the 1D models, and opacity
variations in these shells between the clump and interclump gas mean
than photons can leak through into the surrounding ambient medium. In
the 1D cases, the photoionized region remains trapped in the swept-up
shell.

The results for the 3.5-0.754 2D model without conduction are similar
to their 1D counterpart. The average density inside the hot bubble is
dominated by the long, dense filament that can be seen in
Figure~\ref{fig:PN35_2D} extending inwards from the interface. The
swept-up shell is far broader and has a much lower density than in the
1D models and is partially ionized. In the 2D model with conduction,
the interface is extremely distorted (see Fig.~\ref{fig:PN35_2D_cond})
and this makes it difficult to identify in the averaged density and
temperature plots.

\subsection{Hot bubble mass}

The clumps and filaments formed by the instabilities at the interface
between the hot, shocked fast wind and the dense, photoionized shell
are sources of mass for the hot bubble. Hydrodynamic ablation and
photoevaporation remove material from the dense structures, which then
shocks against and mixes into the hot shocked wind
\citep{Steffen2004}. The mass of hot gas (where we consider $T > 8
\times 10^4$~K to be ``hot'') is thus not solely due to the mass lost
by the post-AGB star. Thermal conduction can also enhance the mass of gas in
the hot bubble by directly evaporating photoionized shell material at
the interface. The increased surface area due to the corrugations in
the interface facilitates the mixing processes.

\begin{figure}
\begin{center}
\includegraphics[width=1.\linewidth]{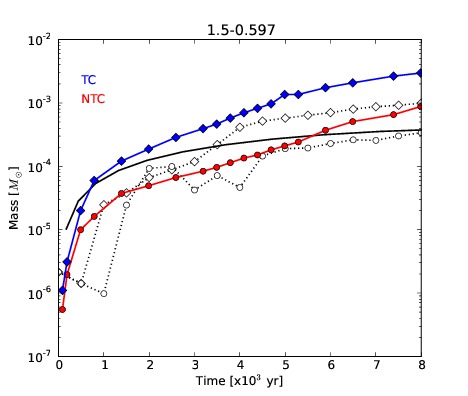}
\includegraphics[width=1.\linewidth]{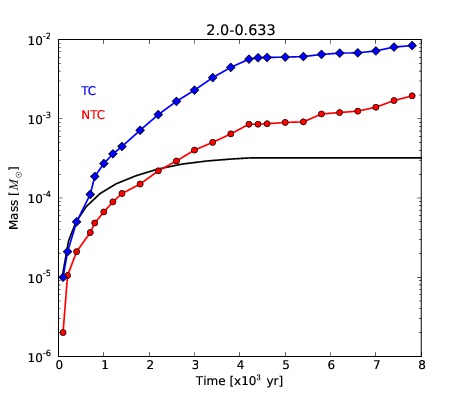}
\caption{The hot bubble mass as a function of time for the 1.5-0.597
  and 2.0-0.633 models. The solid lines without symbols represent the
  amount of mass injected by the fast wind. The lines with circles
  (red) represent the total mass of the hot bubble without thermal
  conduction, while the diamonds (blue) represent the model with
  thermal conduction. Dotted lines with open symbols represent the
  corresponding 1D results for the 1.5-0.597 models only.}
\label{fig:masa_tiempo}
\end{center} 
\end{figure}

To illustrate this, we compute the mass of hot gas in the bubble as a
function of time for the 1.5-0.597 and 2.0-0.633 models (the other
cases are similar) for the first 8000~yr of the
PNe. Figure~\ref{fig:masa_tiempo} shows the total mass injected by the
post-AGB wind, obtained by integrating the mass-loss rates shown in
Figure~\ref{fig:wind_parameters} and the actual mass of hot gas in the
numerical simulation for both models and for the simulations with and
without thermal conduction. It is clear that the hot bubble becomes
dominated by mixed-in mass even in the 2D simulations without
conduction. For the 1.5-0.597 model, this occurs after 5500~yrs when
there is no conduction, and almost from the outset in the model with
conduction. For the 2.0-0.633 model, the mixed-in mass dominates after
2500~yrs in the case without conduction. In contrast, the 1D model for
the 1.5-0.597 case without conduction shows that the hot bubble mass
is always less massive than the total injected wind mass, which is
what one would expect for the case with no mixing. The hot bubble mass
for our 1D model for the 1.5-0.597 case with conduction shows similar
behaviour to the comparable model of Steffen et al (2008), although a
direct comparison is not possible due to the different stellar wind
properties adopted in the two models. Our 2D models show that for the
times when the hot bubble has reached a radius of 0.2~pc, even the
cases without thermal conduction are dominated by mixed-in mass. When
thermal conduction is included, there is an order of magnitude more
mixed-in mass in the hot bubble.

The increase in the mass of hot gas in the PNe in these 2D simulations
as compared to our 1D models (see also the 1D simulations of
\citealp{Steffen2008}) suggests that detection in soft X-rays is
likely even when thermal conduction is not present. We address this
issue in a forthcoming paper (paper II, Toal\'{a} \& Arthur in
preparation).

\subsection{Photoionized shell expansion velocity}

The radial velocity of the photoionized shell is of observational and
theoretical interest
\citep{Pereyra2013,Villaver2002b,Perinotto2004,Schonberner2005b,GarciaSegura2006}
since it provides a link between observable properties of the PN and
the evolution of the central star. The photoionized shell kinematics
depends on the initial density distribution left by the AGB star, the
stellar wind mechanical energy and the stellar ionizing photon
luminosity. Our simulations are focussed on the formation of the hot
bubble (see \S\,2.3), and our spatial domain extends only to 0.5~pc
radius from the star. As a result, at late times we lose information
about the outer shell that results from the ionization front and its
interaction with the outer rim of the AGB material. However, we can
study the expansion velocity of the dense shell created in the fast
wind-AGB-wind interaction. This dense shell, or W-shell
\citep[see][]{Mellema1995}, is created by the expansion of the hot
shocked wind into the photoionized AGB material.

Since the fast wind-AGB-wind interaction provokes the instabilities,
we include them as part of the W-shell for the velocity
calculation. We define the velocity as the density-weighted mean
radial velocity of the photoionized gas in the W-shell by
\begin{equation}
\bar{v}=\frac{\int n_\mathrm{i} v dV}{\int n_\mathrm{i} dV} \ ,
\end{equation}
where the integrals are performed over all photoionized gas with a
ionized number density in excess of 200~cm$^{-3}$.

\begin{figure}
\begin{center}
\includegraphics[width=1.\linewidth]{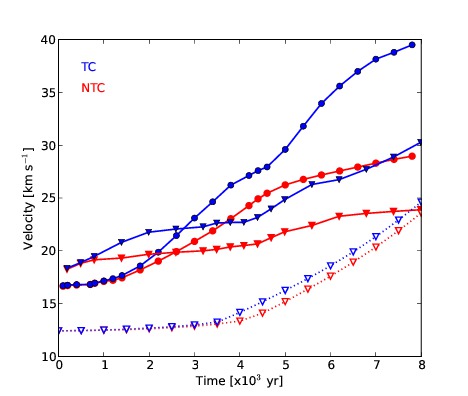}
\caption{W-shell velocity as a function of time for the 1.5-0.597
  (triangles) and 2.0-0.633 (circles) models. Models including thermal
  conduction are represented with filled (red) markers. The
  corresponding 1D results for the 1.5-0.597 model are shown with
  dotted lines and open triangles.}
\label{fig:velocity_ionized}
\end{center} 
\end{figure}

Figure~\ref{fig:velocity_ionized} shows the W-shell velocity as a
function of time for the 1.5-0.597 and 2.0-0.633 models with and
without thermal conduction. All models show a steady increase of the
W-shell velocity with time, which is consistent with previous
numerical results by \citet{Schonberner2005b}. Observational results
\citep{Pereyra2013} suggest that the nebular shells decelerate as the
central star luminosity decreases, however, we do not find this from
our 2D models. Our models with thermal conduction have, in general,
faster expansion rates, since the pressure in the hot bubble is higher
in these models than in those without conduction. We do not calculate
the nebular expansion for the 3.5-0.754 and 5.0-0.90 models as these
would not be observationally detected since they are ionization
bounded and enshrouded by dense neutral material.

\section{Discussion}
\label{sec:discussion}
The results presented in \S\ref{sec:results} illustrate the formation
of hot bubbles for different initial-mass stellar-evolution
models. Instabilities corrugate the interface between the hot bubble
and the surrounding dense shell and lead to the formation of filaments
and clumps. The properties of these clumps depend on the interplay
between the initial AGB density profile and the time-dependence of the
stellar wind and ionizing photon rate in the post-AGB phase. The
stellar wind velocity increases steadily with time for CSPN masses $<
0.754 M_\odot$), and so the temperature of the hot, shocked stellar
wind increases during the PN lifetime. In the case of the
1.0-0.569~$M_\odot$ model, the velocity increases extremely slowly,
and the temperature in the hot bubble is only $T \sim 10^6$~K even
after 5000~yrs of PN evolution. On the other hand, the velocity
increase in the 2.0-0.633~$M_\odot$ model is much more rapid, and
after 5000~yrs, the hot shocked wind temperature is $T > 10^8$~K. For
the most massive star models (3.5-0.754 and 5.0-0.90), the velocity
almost instantaneously increases to $V_w > 6000$~km~s$^{-1}$ but after
only $\sim 1000$~yrs the mechanical energy of post-AGB wind decays
dramatically and the bubble starts to collapse.

Rayleigh-Taylor instabilities are important at the contact
discontinuity between the hot bubble and the dense shell, due to the
continued acceleration of the stellar wind, together with expansion in
a radially decreasing density distribution.  The swept-up AGB material
also cools rapidly and forms a thin, dense shell which is unstable to
thin shell instabilities \citep{Vishniac1983}. Once dense clumps have
begun to form, the shadowing instability \citep{Williams1999} comes
into play due to the variations in opacity, and the density contrasts
between clump and interclump regions can become even more enhanced. If
the density becomes high enough, the clumps become opaque to
photoionizing radiation and the gas within them can recombine.  We
speculate that neutral clumps such as those formed in our 2.5-0.677
simulations could be precursors of molecular clumps observed in PNe
such as NGC\,6053 and NGC\,6720 \citep{Lupu2006,vanHoof2010}. However,
in the present simulations our treatment of the neutral gas is not
sufficiently detailed to follow the formation of molecules and dust.

Another important factor in the formation of the clumps and filaments
is the cooling curve used (see Fig~\ref{fig:cooling}), which for
photoionized regions depends on the abundances in the gas and on the
ionization parameter ($\phi/n$). Our cooling curve was calibrated
using the Cloudy photoionization code for standard PN abundances (see
Table~\ref{tab:abundances_cloudy}), mean stellar effective temperature
$T_\mathrm{eff} = 10^5$~K, gas density $10^3$~cm$^{-3}$ and ionizing
photon flux $\phi = 10^{11}$~cm$^{-2}$s$^{-1}$. There are differences
between results obtained using this cooling curve and cooling curves
for different abundance sets and ionization parameters.

\begin{figure}
\begin{center}
\includegraphics[width=1.\linewidth]{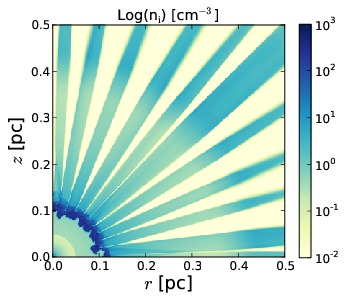}\\
\includegraphics[width=1.\linewidth]{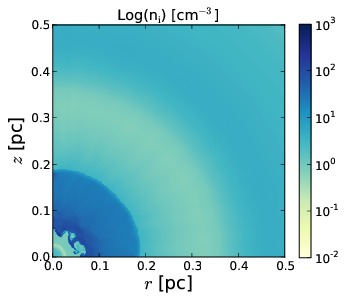}
\caption{Ionized number density for the 1.0-0.569
  model without thermal conduction at 5200~yr. Upper panel:
  simulation using the planetary nebula cooling curve described in
  \S~\ref{sec:num-method}. Lower panel: simulation using
a cooling curve for  standard ISM abundances, stellar effective
temperature $T_\mathrm{eff} = 40000$~K and ionizing flux  $\phi =
10^9$~cm$^{-2}$s$^{-1}$ (see Fig.~\ref{fig:cooling}).} 
\label{fig:cooling_discussion}
\end{center} 
\end{figure}

As an illustration, in Figure~\ref{fig:cooling_discussion} we compare
the total ionized number density resulting from a simulation for a
1.0-0.569 model without thermal conduction using our usual PN cooling
curve with that of a simulation using a cooling curve calibrated for a
cooler star ($T_{\mathrm{eff}}=40$~kK) with standard ISM abundances
taken from Cloudy (see Table~\ref{tab:abundances_cloudy}), gas density
1~cm$^{-3}$ and ionizing photon flux $\phi = 10^9$~cm$^{-2}$s$^{-1}$
(Fig.~\ref{fig:cooling}--thin-solid line). The appearance of the
interface is very different: there are fewer clumps and the
instability appears to be more a pure Rayleigh-Taylor type. The
density of the clumps at the interface is not large enough to trap the
ionizing photons and so the region beyond the swept-up shell is
completely photoionized. The differences due to the adopted cooling
curves are most important in the temperature range $10^{4} \mathrm{K}
< T < 10^{6} \mathrm{K}$. Cooling below 10$^{5}$~K depends on stellar
parameters such as effective temperature and ionizing photon rate,
while cooling above 10$^{5}$~K is independent of these factors, since
the hot gas is collisionally ionized, although the metallicity remains
an important ingredient \citep[see][]{Steffen2012}. The cooling curve
used in the present simulation is a clear improvement over the
standard ISM CIE cooling curve typically used in hydrodynamical
simulations because it uses a more appropiate abundance set and takes
into account the ionizing photon flux from the central star, but
tailoring the stallar parameters, ionization parameter, and abundance
set at each time would be more accurate. A full investigation of this
effect is beyond the scope of the present work. Finally, our cooling
assumes ionization equilibrium but cooling can also be modified if
nonequilibrium ionization is taken into account (see
\citealp{Steffen2008}).

The clumps and filaments are sources of mixing due to hydrodynamic
ablation, shock diffraction and photoevaporation and the large surface
area leads to substantial quantities of dense material being shocked
and incorporated into the hot bubble. For models without thermal
conduction, the breakup of the dense shell can cause hot gas to leak
out of the bubble, flowing between the clumps and filaments and
thereby depressurising the hot bubble. The inclusion of thermal
conduction leads to evaporation of material at the interface, which
fills in the interclump region in the broken shell and reseals the
bubble. This results in higher pressures in the conduction models than
in those without conduction. Indeed, the expansion of the models
without thermal conduction is dominated by the pressure of the thick
shell of photoionized gas around the hot bubble, while the models with
thermal conduction are more similar to wind-blown bubbles, surrounded
by thin shells of swept-up photoionized material.

We have investigated the extent to which numerical diffusion is
responsible for mixing across the interface by running a test
simulation consisting of an inner, circular region of hot, diffuse gas
in pressure equilibrium with a surrounding envelope of dense, cool
gas. Both gases were assumed to be at rest, but radiative cooling was
taken into account. After 10,000~yrs, the interface between the two
fluids was only 4 cells wide and was not growing with time. This gives
us confidence that the mixing seen in our simulations without
conduction is a real, hydrodynamical effect rather than a numerical
one. We have also undertaken studies of the effect of increasing and
decreasing the numerical resolution and the size of the wind injection
region. The qualitative results are the same, although small details
do vary.

\subsection{Comparison with previous works}

There are some differences that are worth noting between our numerical
calculations and the referenced works mentioned in
\S~\ref{sec:intro}. We limit our comparison to those works that have
taken into account a detailed treatment of the evolution of the
stellar wind parameters and show the formation of hot bubbles, principally
those studies presented by \citet{Villaver2002a,Villaver2002b},
\citet{Perinotto2004} and \citet{Steffen2008}.

Potential differences are due to the stellar evolution models, which
determine both the initial conditions defined by the final few thermal
pulses of the AGB state and the stellar wind parameters in the
post-AGB phase, the numerical method and the physics included in the
model.

We adopt the same AGB wind parameters as \citep{Villaver2002a}, i.e.,
those calculated by \citet{Vassiliadis1993} and so do not expect large
departures from their results. Indeed, the density profiles we arrive
at at the end of the TP-AGB stage are very similar, particularly the
innermost 0.5~pc, which is of most interest to us. We find steep
density profiles ($\rho \propto r^{<-2}$) for low initial stellar mass
models and shallow density profiles ($\rho \propto r^{>-2}$) for the
high initial stellar mass models.  \citet{Perinotto2004} and
\citet{Steffen2008} consider two types of AGB winds: a constant
mass-loss rate and wind velocity model, which gives a $\rho \propto
r^{-2}$ density profile, and a ``hydrodynamical'' model, where the
mass-loss rate and wind velocity are based on the AGB mass-loss
prescription of \citet{Blocker1995}, which leads to a variable density
gradient. As pointed out in these previous works, once the
photoionization switches on, the fine details of the initial density
and velocity profiles are quickly erased by the shock that passes
through the neutral gas ahead of the ionization front.

For the post-AGB evolution, we use the same stellar evolution models
as \citet{Villaver2002b}. However, we calculate the stellar wind mass
loss and terminal velocity using the WM-Basic code \citep[][and
subsequent papers.]{Pauldrach1986}. We also obtained the ionizing
photon flux from WM-Basic instead of using a black body. Our stellar
wind parameters differ from those of \citet{Villaver2002b}, who use
the empirical fits of \citet{Vassiliadis1994} to the data compiled by
\citet{Pauldrach1988}. The resulting mass-loss rates are in the same
range of values with very similar temporal dependence. On the other
hand, the terminal wind velocities obtained by \citet{Villaver2002b}
are a factor of approximately 2 higher than ours for all stellar
models presented in our Figure~\ref{fig:wind_parameters} (top-right
panel).

\citet{Perinotto2004} and \citet{Steffen2008} use different
evolutionary models for the post-AGB stage
\citep{Blocker1995,Schonberner1983}. Moreover, for stellar effective
temperatures below 25~kK they use the \citet{Reimers1975} prescription
for the stellar wind mass-loss rate. Their stellar wind properties are
therefore not quite the same as those used in this paper---our
mass-loss rates are slightly higher and our wind terminal velocity is
slightly lower---but the differences are not significant since the
stellar wind mechanical luminosity $L_\mathrm{w} = 0.5 \dot{M}
V_\mathrm{w}^2$ is the important quantity. Once again, a black body
spectrum is used to calculate the ionizing photon rate.

The minor differences in the stellar wind parameters and ionizing
photon rate between our models and previously published work are
dynamically not that important. The duration of the transition from
the AGB to the white dwarf stage is the main indicator of whether the
planetary nebula will be density bounded (as seen for the models with
initial mass $M < 2.5 M_\odot$) or ionization bounded (as seen for the
3.5-0.754 and 5.0-0.90 models). Our results agree broadly with those
of \citet{Villaver2002b} and \citet{Perinotto2004} in this regard but
the fact that our simulations are 2D allows the possibility that the
shell of AGB material swept up by the fast wind in the early stages of
the PN evolution breaks up due to instabilities. For the lower initial
stellar mass models, this creates regions of higher opacity (clumps)
that may even recombine and become neutral condensations embedded in a
surrounding photoionized flow, and regions of lower opacity through
which the ionizing photons may pass to photoionize the material beyond
the shell. The formation of these clumps starts with the thin shell
instability and depends on the metallicity of the gas and on the form
of the cooling curve. The evolution of the photoionized gas within and
beyond the shell depends on the radiative transfer model. Our code
includes a self-consistent treatment of the radiative transfer and we
are able to follow regions of partially ionized, recombining gas
\citep[see also ][]{Perinotto2004,Steffen2008}, in contrast to
\citet{Villaver2002b} who use the Str\"omgren approximation.

Unlike \citet{Steffen2008}, we find that in our 2D models the thermal
conduction does have a dynamical effect on the planetary nebula
system. For the lower mass CSPN ($M_{\mathrm{WD}}<0.677M_\odot$),
models without conduction have leaky bubbles, whereas thermal
conduction reseals the gaps between the clumps and filaments in the
broken shell. The gas between the clumps and filaments is material
which has been evaporated from the filaments and has expanded as it is
heated. The hot bubble in these models grows much more quickly,
reaching our ``target'' distance of 0.2~pc much sooner than in the
models without conduction, since there is no loss of pressure, and the
interclump material has temperatures $T \sim 10^5$~K. The interclump
material in the models without conduction, on the other hand, is hot
gas with $T > 10^6$~K escaping from the bubble.

Figure~\ref{fig:compare1d2d} shows that for the higher mass models,
the positions of the inner and outer shock waves and the interface
between the hot bubble and swept-up shell for models both with and
without conduction for the 2D simulations averaged over angle coincide
reasonably well with their 1D counterparts. However, in the 2D cases,
a substantial amount of dense shell material is shed from the clumps
and filaments by hydrodynamic interaction, photoevaporation and
thermal evaporation and is thoroughly mixed throughout the hot bubble
by non-radial motions. At later times, once the dynamical time has
become as large as the cooling time, cooling will become important
inside the hot bubble. This will occur much earlier in the 2D
simulations than in the 1D simulations because the density is much
higher in the former.

While cooling is still unimportant, the effect of thermal conduction
is to raise the temperature in the cool, dense gas at the interface
with the hot bubble. In both 1D and 2D simulations, the heated gas
expands and pushes the swept-up shell outwards. This is particularly
the case with the higher-mass CSPN models
$M_{\mathrm{WD}}>0.677M_\odot$). These models have larger hot bubbles
than their counterparts without conduction, although they still remain
enshrouded by neutral material and would not be detectable.

Our 2.5-0.677 model shows intermediate behaviour. The simulation
without conduction has ``leaky bubble'' characteristics, while the
simulation with conduction mixes quite a lot of dense shell material
back into the bubble. In these models, the clumps in the broken shell
recombine and photoevaporated material flows off the heads of these
clumps back towards the star.

The effect of thermal conduction on the dynamics is also evident in
Figure~\ref{fig:velocity_ionized}, where we show the W-shell (dense
shell formed by the fast wind-AGB wind interaction) velocity as a
function of time for two of our lower mass CSPN models, 1.5-0.597 and
2.0-0.633. For these models, the simulations without conduction are of
the ``leaky bubble'' type, where hot, shocked gas pushes its way
between the clumps and filaments of the broken shell. The reduced
pressure in the hot bubble results in lower expansion velocities for
the external photoionized shell. The corresponding models with
conduction remain pressurised, since the gaps in the shell are sealed
by material evaporated from the filaments. The expansion velocities in
the surrounding photoionized shell are correspondingly higher.

\citet{MellemaFrank1995} and \citet{Mellema1995} performed 2D
axisymmetric radiation-hydrodynamics simulations of the formation of
aspherical planetary nebulae through an interacting winds
model. Their models took into account the density distribution of the
circumstellar AGB stage material and although \citet{MellemaFrank1995}
considers constant fast wind parameters, \citet{Mellema1995} takes
into account the time evolution of the post-AGB wind and ionizing
photons. Unfortunately, the resolution of these simulations is very
low (80x80 computational cells at best), and although the formation of
instabilities at the interface between the hot shocked fast wind and
the swept-up photoionized AGB material is hinted at, the numerical
resolution is simply not adequate to follow their
development. \citet{MellemaFrank1995} estimate the soft X-ray emission
from their models and find it to come from a thin layer just inside
the photoionized shell. However, this is an artefact due to numerical
diffusion at the contact discontinuity and the low resolution of the
simulations converts it into an appreciably amount of soft X-ray
emission. Other 2D, purely hydrodynamic simulations by
\citet{Stute2006} show the formation of instabilities at the contact
discontinuity, but these remain confined to a thin layer and do not
grow. In our simulations, the resolution is much higher and the
instabilities can be characterised and are well developed. Moreover,
the strong radiative cooling and the proper treatment of the radiative
transfer in our models mean that once the instabilities form they
become enhanced and grow in size so that the surface area they present
becomes quite large.

We only follow the evolution of our models up until the hot bubble has
achieved a radius of $\sim 0.2$~pc. For the 1.0-0.569, 1.5-0.597, and
2.0-0.633 models this occurs before the mass-loss rate and ionizing
photon luminosity have dropped off completely and so it can be
expected that these hot bubbles will expand further. The 2.5-0.677
model reaches 0.2~pc only very slowly, since the mass-loss rate and
ionizing photon luminosity decline sharply after 3000~yrs and it is
the residual pressure of the hot bubble that is driving the expansion
against the diminishing pressure of the expanding photoionized
shell. Our most massive star models, 3.5-0.754 and 5.0-0.90, have such
a fast evolution that the hot bubble reaches a maximum radius before
the pressure in the hot shocked gas falls below the pressure of the
surrounding photoionized shell and the hot bubble collapses back
towards the star. This is a natural way of explaining the back-filling
of gas in PNe, such as is seen in the Helix nebula
\citep{{Meaburn2005},{GarciaSegura2006}}.

Finally, we speculate that in the case of 3D simulations the
instabilities in the wind-wind interaction zone would develop faster
than the formation of clumps in 2D simulations
\citep[see][]{Young2001}. This was explored in numerical simulations
for the analogous case of the formation of (hot bubbles in) WR nebulae
by \citet{vanMarle2012}. They found that 3D simulations show more
structure than 2D simulations, although the differences are relatively
small. In 3D, there will be a larger surface area of dense AGB
material available to be hydrodynamically ablated or evaporated and we
therefore expect that the mixing of material into the hot bubble would
be even more effective than in the present 2D results.

\section{Summary and conclusions}
\label{sec:summary}

In this paper we have presented high-resolution 2D
radiation-hydrodynamic numerical simulations of the formation and
evolution of hot bubbles in planetary nebulae (PNe). We considered six
different stellar evolution models with initial masses of 1, 1.5, 2,
2.5, 3.5, and $5 M_{\odot}$ at solar metallicity
\citep[$Z=0.016$;][]{Vassiliadis1993,Vassiliadis1994}, which
correspond to final WD masses of 0.569, 0.597, 0.633, 0.677, 0.754,
and 0.90~$M_{\odot}$. The initial circumstellar conditions are
obtained from 1D simulations of the final few thermal pulses of the
AGB phase and are the same as those used in \citet{Villaver2002a}, but
in the case of the post-AGB phase the stellar wind velocity, mass-loss
rate, and ionizing photon rate are computed self-consistenly using the
WM-Basic stellar atmosphere code \citep[][and references
  therein]{Pauldrach2012}. From our results we find that:

\begin{itemize}

\item The formation and evolution of hot bubbles in PNe depend on
  different factors: the density distribution of the previously
  ejected AGB material, the time-evolution of the stellar wind
  parameters in the post-AGB phase, and the metallicity and cooling
  rate used in the calculations. Differences between models create
  different patterns of instabilities in the wind-wind interaction
  zone. The most important factors are the length of time before the
  mass-loss rate and ionizing photon rate decline sharply, and the
  acceleration timescale for the stellar wind.

\item The breakup of the shell of swept-up AGB material due to
  instabilities complicates the wind-wind interaction region and none
  of our 2D models exhibits a clear double-shell pattern such as those
  observed by \citet{Guerrero1998}. The formation of clumps leads to
  dynamic opacity variations in the photoionized shell, which result
  in bright-rimmed clumps and a ray-like distribution of lower density
  photoionized gas in the AGB envelope, which vary with time.

\item The formation and distribution of clumps and filaments has been
  shown to depend strongly on the cooling rate curve used and the
  ionization and temperature history of these clumps should be a
  function of the evolving stellar parameters and radiation
  field. Although we take into account planetary nebula abundances and
  a representative radiation field when calculating the radiative
  losses, we do not take into account the time evolution of the
  stellar parameters nor do we include molecules or dust in our
  treatment of the neutral gas.

\item The creation of hydrodynamical instabilities results in
  mixing of AGB material into the hot bubble, even for cases in which
  thermal conduction is not included. The temperature and density of
  the mixing zone are favourable for the emission of soft X-rays.

\item The inclusion of thermal conduction has several effects: it
  stops the bubble leaking hot gas through the gaps between the clumps
  and filaments by sealing them with evaporated material, and it
  increases the amount of AGB material mixed into the hot
  bubble. Models without conduction were found to form leaking bubbles
  with pressures dominated by the photoionized shell, whereas models
  with conduction resemble wind-blown bubbles surrounded by thin
  swept-up shells. This changes the appearance of the clumps around
  the edge of the hot bubble in the two cases. Finally, in cases in
  which thermal conduction could be suppressed by a magnetic field,
  hydrodynamical instabilities will dominate the mixing of material
  into the hot bubble.

\end{itemize}

The next paper in this series will consider the synthetic X-ray
emission (i.e., emission measure, luminosity, surface brightness, and
simulated spectra) resulting from these simulations.

\section*{Acknowledgments}

We thank the referee, Matthias Steffen, for valuable
suggestions that improved the presentation of this paper. We would like to
thank E.\,Villaver for helpful discussions and for providing the AGB
mass-loss rate and velocity parameters of the six stellar evolution
models used here. We thank M.A.\,Guerrero for a critical reading of
the manuscript and W.\,Henney for useful comments. We are grateful to
Y.\,Jim\'{e}nez-Teja and K.\,\'{A}lamo-Mart\'{\i}nez for
introducing us to the use of the matplotlib routines. SJA and JAT
acknowledge financial support through PAPIIT project IN101713 from
DGAPA-UNAM. JAT thanks support by CSIC JAE-Pre student grant
2011-00189.

\appendix
\section{Additional Figures}
\label{app:appa}
In this appendix we collect together figures for the models not
presented in the main text.

\begin{figure*}
\includegraphics[width=1.0\linewidth]{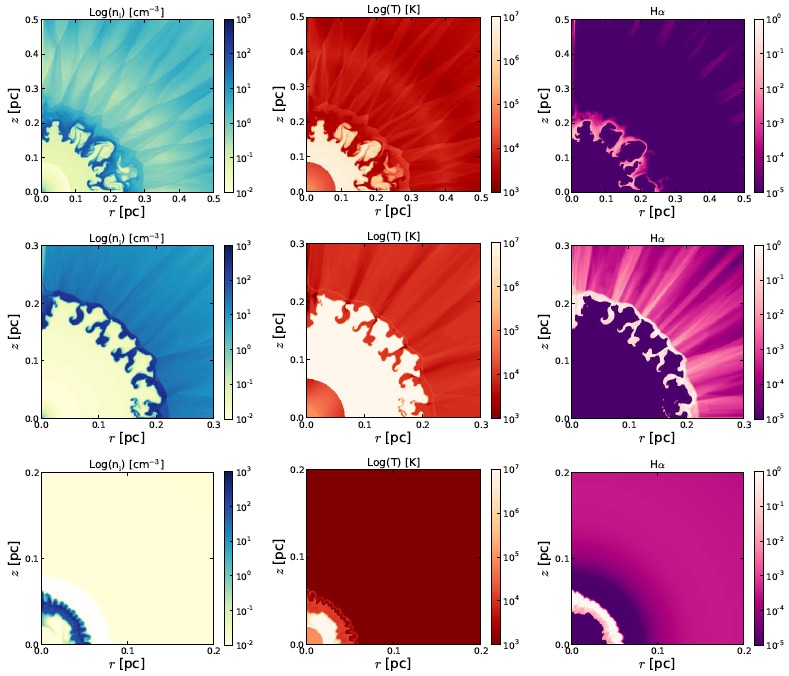}
\caption{Additional models without conduction. Top row: 1.0-0.569
  model at 10,200~yrs (when $T_{\mathrm{eff}}=31260$~K and $L=3615
  L_{\odot}$ for this model). Middle row: 2.0-0.633 model at 4200~yrs
  (when $T_{\mathrm{eff}}=171790$~K and $L=3200 L_{\odot}$ for this
  model).  Bottom row: 5.0-0.90 model at 770~yrs (when
  $T_{\mathrm{eff}}=213800$~K and $L=400 L_{\odot}$ for this
  model). The first and second rows correspond to the hot bubble
  having a mean radius of 0.2~pc. The third row corresponds to the
  maximum radius of the hot bubble for the 5.0-0.90 model. From left
  to right, the panels represent: total ionized number density (left
  panels), temperature (middle panels), and normalized H$\alpha$
  volume emissivity (right panels).}
\label{fig:PNall_2D}
\end{figure*}

\begin{figure*}
\includegraphics[width=1.0\linewidth]{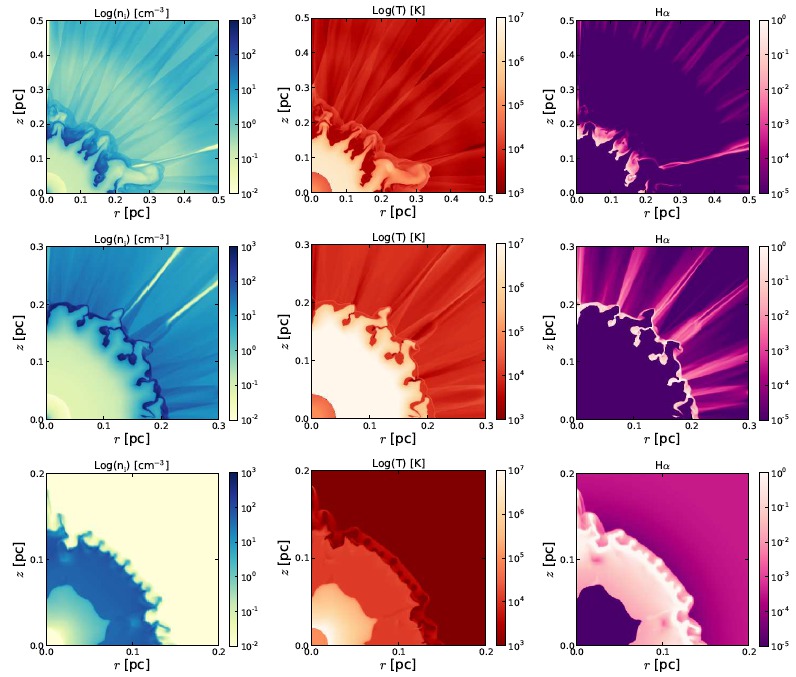}
\caption{Additional models with thermal conduction. Top row: 1.0-0.569
  model at 8700~yrs (when $T_{\mathrm{eff}}=29040$~K and $L=3622
  L_{\odot}$ for this model). Middle row: 2.0-0.633 model at 3600~yrs
  (when $T_{\mathrm{eff}}=159220$~K and $L=4520 L_{\odot}$ for this
  model).  Bottom row: 5.0-0.90 model at 3370~yrs (when
  $T_{\mathrm{eff}}=180720$~K and $L=180 L_{\odot}$ for this
  model). The first and second rows correspond to the hot bubble
  having a mean radius of 0.2~pc. The third row corresponds to the
  maximum radius of the hot bubble for the 5.0-0.90 model. From left
  to right, the panels represent: total ionized number density (left
  panels), temperature (middle panels), and normalized H$\alpha$
  volume emissivity (right panels).}
\label{fig:PNall_2D_cond}
\end{figure*}

Figure~\ref{fig:PNall_2D} presents final time results for the
1.0-0.569, 2.0-0.638, and 5.0-0.90 models without conduction. For the
lower mass models, we define the final time as when the mean radius of
the hot bubble reaches 0.2~pc.  For the high mass model (5.0-0.90) the
final time corresponds to the maximum radius of the hot bubble, after
which time it begins to collapse.  Figure~\ref{fig:PNall_2D_cond}
presents final time results for the same models with thermal
conduction. Models 1.0-0.569 and 2.0-0.638 belong to Group~A and
exhibit similar behaviour to the 1.5-0.597 model described in the main
text. Model 5.0-0.90 is a Group~B model and behaves in a similar
fashion to the 3.5-0.754 model presented in the main text.

\end{document}